\documentclass[a4paper,11pt]{article}
\usepackage{amsfonts}

\usepackage{graphicx}
\usepackage{amsmath}
\usepackage{amssymb}
\usepackage{latexsym}
\usepackage[colorlinks]{hyperref}
\usepackage{color}
\usepackage{float}
\usepackage{cite}
\usepackage{makeidx}
\usepackage{colortbl}
\usepackage{csquotes}
\usepackage[squaren]{SIunits}

\usepackage[textfont={footnotesize,sf},labelfont={color=blue,bf,sf},labelsep=endash]{caption}
\usepackage[position=top,labelfont={color=blue,bf,sf}]{subfig}
\usepackage{bm}

cm \voffset = -2truecm

\newcommand{\bra}{\begin{array}}
\newcommand{\era}{\end{array}}
\newcommand{\beq}{\begin{equation}}
\newcommand{\eeq}{\end{equation}}
\newcommand{\beqar}{\begin{eqnarray}}
\newcommand{\eeqar}{\end{eqnarray}}

\def\BC{\bb C}
\def\_\BC{\bbi C}

\def\( {\left(}
   \def\) {\right)}
\def\[ {\left[}
\def\] {\right]}
\def\no2 {{\textstyle{n\over 2}}}

\newcommand{\lga}{\longrightarrow}

\newcommand{\lb}{\label}

\begin{document}
\begin{titlepage}
\setcounter{page}{1}
\renewcommand{\thefootnote}{\fnsymbol{footnote}}
\begin{flushright}
\end{flushright}
\begin{center}
{\Large \bf {Energy Levels of Graphene  Magnetic Circular
Quantum Dot
}}

\vspace{5mm}

{\bf Abdelhadi Belouad}$^a$, {\bf Bouchaib Lemaalem}$^a$,  {\bf
Ahmed Jellal$^a$\footnote{\sf 
a.jellal@ucd.ac.ma}} and {\bf Hocine Bahlouli}$^b$

\vspace{5mm}

{$^a$\em Laboratory of Theoretical Physics,  
Faculty of Sciences, Choua\"ib Doukkali University},\\
{\em PO Box 20, 24000 El Jadida, Morocco}



{$^b$\em Physics Department,  King Fahd University
of Petroleum $\&$ Minerals,\\
Dhahran 31261, Saudi Arabia}


\vspace{3cm}

\begin{abstract}
We study the energy levels
of graphene magnetic circular quantum dot surrounded by an infinite  graphene
sheet in the presence of an electrostatic potential. We solve Dirac equation to derive the solutions of energy spectrum
associated with different regions composing our system.
 Using the continuum model and applying boundary conditions at the interface, we obtain analytical results for the energy levels. The dependence of the energy levels 
 on the quantum dot radius, magnetic
 field and electrostatic potential is analyzed for the two valleys $K$ and $K'$. We show
 that the energy levels exhibit characteristics of interface states and
 have an energy gap.

\end{abstract}
\end{center}

\vspace{3cm}



\end{titlepage}

\section{Introduction}

Graphene has been the
subject of massive research throughout the world since the first experiments in
2004 \cite{Novoselov04, {Novoselov06}} 
because of its unique electronic properties that could be important for nanoelectronics applications
\cite{Zhang05,Chung10,Choe10}. Graphene was prepared using several techniques, including silicon carbide surface precipitation \cite{Berger06, Ohta06}, mechanical exfoliation from graphite 
and chemical vapor deposition growth on a catalytic metal surfaces \cite{Kim09, Reina08, De Arco09}. The electronic structure of graphene is well described elsewhere \cite{Castro09}
involving two nodal zero-gap
points $(K, K')$, called Dirac points, in the first Brillouin zone at which the conduction and valence
bands touch.
That leads to a number of its unusual peculiar electronic properties such as linear dispersion relation, gapless {energy spectrum} and so on
\cite{Castro09,Geim09, Geim07}.

Graphene quantum dots (QDs) are small graphene fragments, where electronic {wavefunction} is confined in {disk of radius $R$}. 
Excitons in graphene have an infinite Bohr diameter. Thus, graphene fragments of any size will show quantum confinement
effects. As a result, graphene QDs have a non-zero bandgap and
luminescence on excitation. This bandgap is tunable by modifying
the size and surface chemistry of the graphene QDs.
After its discovery, researchers have attempted to confine electrons in
graphene-based QDs because of the wide range of new applications of QDs for example in electronic circuits, photovoltaic systems \cite{Bacon13}, qubits \cite{Trauzettel07} and gas detection \cite{Sun13}. Graphene as the basis of these QDs could enable fast and flexible devices. In general, the ultra-relativistic nature of graphene charge carriers has led researchers to wonder how they would react to confinement \cite{Rozhkov11}. However, it is precisely this particular property that prohibits the use of traditional manufacturing techniques such as local electrostatic {bias} to confine carriers. The Klein tunneling effect \cite{Katsnelson06} allows electrons to use hole states in the gated region to escape the QD. For instance, one has tried using magnetic fields \cite{Espinosa13, Martino07}, cutting the flake into small nanostructures \cite {Mirzakhani16,Zebrowski13} or using the substrate to induce a band gap \cite{Recher09}. However, magnetic fields bring along many difficulties in nano-sized systems \cite{Liu17}. The question of confining Dirac Fermions in graphene QDs has resulted in many propositions. QDs made from nanostructures are highly sensitive to the precise shape of the edge, which is hard to control \cite{Espinosa13}.

{
To complete our literature review we would like to mention previous work on graphene QDs. The 
circularly symmetric graphene nanostructures in the form of graphene rings, dots, and antidots were studied 
by demonstrating an excellent agreement with atomistic models for small structures \cite{Thomsen}. Experimentally, 
electrostatically confined monolayer graphene QDs with
orbital and valley splittings was realized \cite{Freitag2}. 
%
{The hybrid monolayer-bilayer graphene QDs were investigated in \cite{Mirzakhani16} by considering
a circular single-layer graphene quantum dot surrounded by an infinite bilayer graphene sheet as well as a circular bilayer graphene quantum dot surrounded by an infinite single-layer graphene.}
%
Here the QD  boundary conditions (zigzag or armchair) between the QD and its surrounding infinite graphene medium are very important, the QD energy levels exhibit interface states characteristics. Unlike our system (monolayer magnetic QD embedded in an infinite graphene sheet) the external electrostatic field can induce a tunable energy gap in the energy spectrum in such case. 
{By solving Dirac equation an analytical solution to calculate energy levels and
wave functions of mono- and bilayer graphene quantum
dots was presented in
\cite{Tamandani}. All these important works are different from the present work but some common features of graphene QD will be pointed out in the conclusion, such as the decrease of the energy band gap as the size of the QD increases.
}

We study the confinement of the charge carriers in a
magnetic circular quantum dot in graphene  surrounded by an infinite
graphene sheet.
{
In our case we are dealing with a graphene magnetic circular quantum dot who boundaries are defined by the profile of the applied magnetic field, hence we do not have terminated boundaries that result in dangling bonds. Hence 
zigzag and armchair boundary conditions are not relevant to our situation, contrary to previous works \cite{Mirzakhani16, Tamandani}.
}
The corresponding band structures will be analyzed by deriving
the solutions of the energy spectrum.
%
Subsequently, by applying the boundary condition at the interface we obtain
an equation describing the
energy levels in terms of the physical parameters characterizing
our system. We numerically investigate
the dependence of such energy levels
on quantum dot radius, magnetic field and electrostatic potential.
In particular, we show that the energy levels exhibit different
symmetries and energy gap under various conditions.
Indeed, we show that the energy levels are degenerate in the case where
the QD radius goes to zero ($R\lga 0$). However as long as $R$ increases
we obtain two set of energies showing the symmetric and asymmetric behaviors.
In addition, under some conditions we show that the electron density
can be modified by the presence of electrostatic potential.

The present paper is organized as follows. In section $2$, we solve Dirac equation to obtain the eigenspinors describing fermions in graphene
magnetic quantum dot surrounded by an  infinite graphene sheet.
These solutions will be used together with the continuity condition
to determine the corresponding energy levels.
%
We numerically discuss the energy spectrum and the electron density under various choices
of the physical parameters in section 3.
We conclude our results
in the final section.

%


\section{Model and theory}

We consider  a graphene based quantum dot of radius $R$ with  magnetic circular geometry   surrounded from exterior with an infinite graphene sheet, schematically depicted in Figure \ref{f00}.
The dynamics of carriers in the
honeycomb lattice of covalent-bond carbon atoms of single
layer graphene can be described by the Hamiltonian
\begin{equation}\label{e1}
H=v_F\vec \sigma \cdot \left(\vec p+e \vec A\right)+U(r) \mathbb{I}
\end{equation}
where $v_F = 10^6$ m/s is the Fermi velocity, $p=(p_x,p_y)$ is
the two-dimensional momentum operator,
$\sigma=(\sigma_x,\sigma_y)$ are Pauli spin matrices in the
basis of the two sublattices of $A$ and $B$ atoms, $U(r)$ is an axially symmetric electrostatic potential applied to the system, $\overrightarrow{A}$ is the vector potential in the symmetric gauge and $\mathbb{I}$ is the $2\times2$ identity matrix. 

\begin{figure}[h!]
  \center
  \includegraphics[width=8cm]{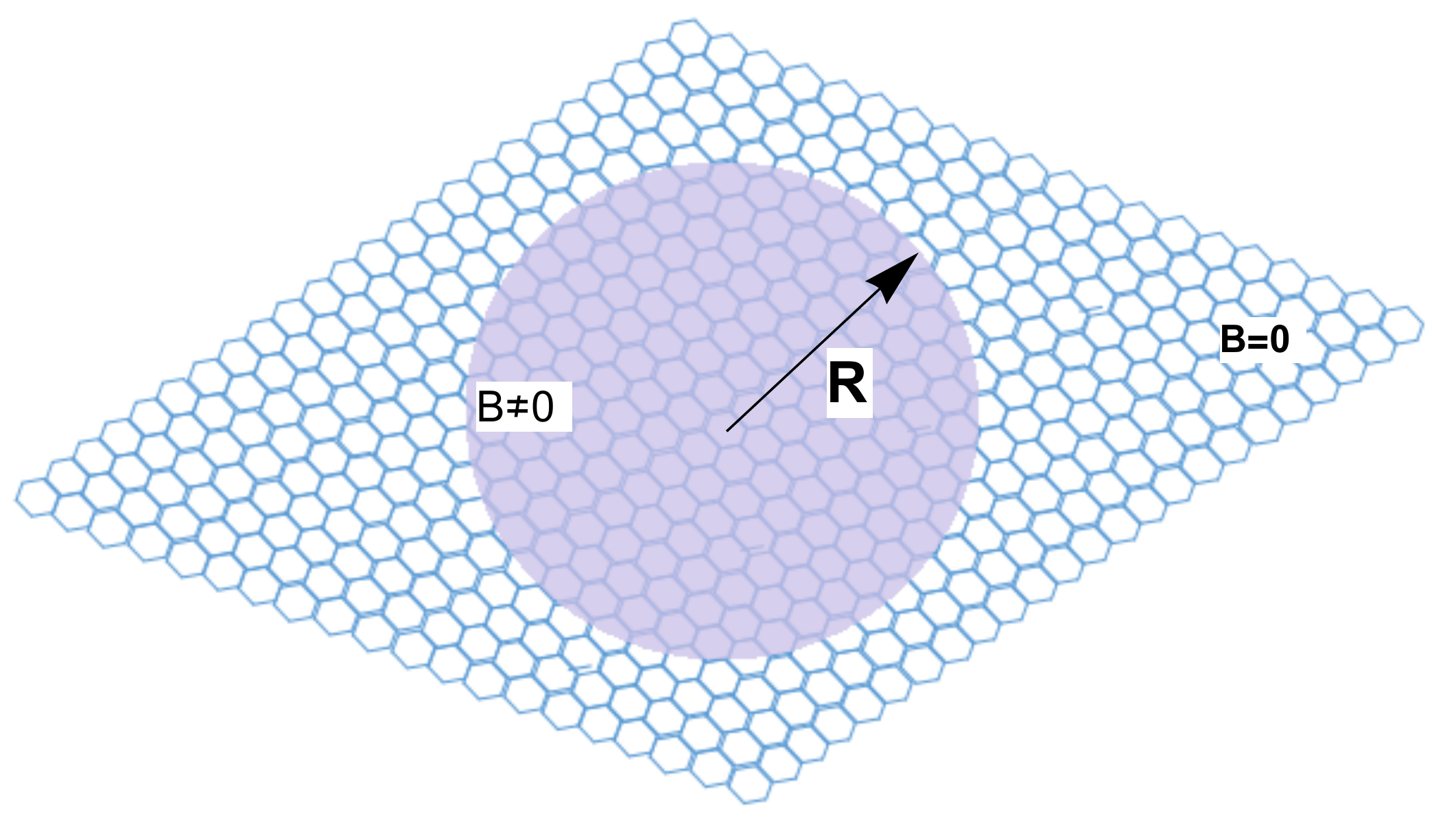}
  \caption{\sf (color online) Schematic diagram
  of graphene magnetic circular Quantum dot of radius $R$ surrounded by an infinite graphene sheet in the presence of perpendicular magnetic field $B$ inside the quantum dot.
  \label{f00}}
\end{figure}

We assume
that the carriers are confined in a circular area of radius $R$.
 Due to the circular symmetry we have $[H,J_z]=0$ with $J_z=-i\hbar\partial_\varphi+\hbar\sigma_z/2$ is the total angular momentum. This implies that the two component
wavefunctions
in the polar coordinates $(r,\varphi)$ take the form
\begin{equation}\label{e2}
\Psi^\tau(r,\varphi)=
e^{im\varphi}\left(%
\begin{array}{c}
  \varphi^\tau_A(r) \\
  ie^{-i\tau\varphi}\varphi^\tau_B(r) \\
\end{array}%
\right)
\end{equation}
where $m=0,\pm1,\pm2 \cdots $ being the orbital angular momentum
quantum number. The radial components $ \varphi^\tau_A$ and $
\varphi^\tau_B$ express amplitude probabilities on the two carbon
sublattices of graphene,
  the quantum number
$\tau=\pm 1$ distinguishes the two  valleys $K$ and $K'$.
{
It is worth mentioning that the low energy effective theory of graphene results in two equivalent or degenerate valleys that contribute equally to the transport properties. These valleys are located at two inequivalent points at the edge of the Brillouin zone called $K$ and $K'$. Whenever an external applied field couples differently to the carriers in the two valleys, then the valley degeneracy is lifted and interesting physical observation might be realized. The mathematical distinction between these two valleys is easily associated with a valley tag or quantum number that takes the value of $+1$ and $-1$ in these respective valleys. Different proposal have been advanced to create such a valley polarization in graphene. The main ingredient in application would be to control the single valley carrier occupancy in graphene, hence resulting in the so-called valley polarization. So much in the same way as the electron spin is used in spintronics or quantum computing, the valley quantum number can be used in "valleytronics", the valley quantum number so to speak has also two values +1 and -1 \cite{group}.
}

To discuss the localized-state solutions in the present system, we consider
a circularly symmetric QD
subject to  the following magnetic field
\begin{equation}\label{e3}
\vec{B}=
\left\{
  \begin{array}{ll}
    B\vec{e}_z, & \hbox{$r<R$} \\
    0, & \hbox{$r>R$}
  \end{array}
\right.
\end{equation}
and due to continuity, the corresponding vector potential reads 
\begin{equation}\label{e4}
\vec A=
\left\{
  \begin{array}{ll}
    \frac{B}{2r}(r^2-R^2) {\vec e_\varphi}, & \hbox{$r<R$} \\
    0, & \hbox{$r>R$}
  \end{array}
\right.
\end{equation}
{and $\vec e_\varphi$ is the unitary vector.}
To go further in obtaining the solutions of the energy spectrum, one has to distinguish two different cases.
Indeed for $r>R$ (absence of magnetic field) the Hamiltonian \eqref{e1} reduces
to the following
\begin{equation}\label{e5}
H=\left(%
\begin{array}{cc}
  U & \pi_+ \\
  \pi_- & U \\
\end{array}%
\right)
\end{equation}
where the momentum operators are given by
\beq\label{e6}
\pi^{\pm}=-i\hbar v_Fe^{\pm i\tau\varphi}\left(\frac{\partial}
{\partial r}\pm\frac{\tau i}{r}\frac{\partial}{\partial
\varphi}\right).
\eeq
{Using the eigenvalue equation  $H\Psi_{>}^\tau=E\Psi_{>}^\tau$ 
along with the two component wavefunctions \eqref{e2}}
we obtain
\begin{eqnarray}\label{e8}
&&
  \left(\frac{\partial}{\partial \rho}+\frac{m\tau}{\rho}\right)\varphi^\tau_{A}(\rho)=-(\varepsilon-u)\varphi^\tau_{B}(\rho)\\
  &&
  \label{e88}
  \left(\frac{\partial}{\partial \rho}+\frac{m\tau-1}{\rho}\right)\varphi^\tau_{B}(\rho)=(\varepsilon-u)\varphi^\tau_{A}(\rho)
\end{eqnarray}
{where all energies are measured in units of $E_0=\frac{\hbar v_F}{R}$
and
 dimensionless units $\rho=\frac{r}{R}$,  $\varepsilon=\frac{E}{E_0}$, $u=\frac{U}{E_0}R$ have been introduced, 
  the subscript $>$ means that this solution holds for $r>R$}.
Now injecting  \eqref{e8} into \eqref{e88} to get the second order
differential equation for $\varphi^\tau_{A}(\rho)$
\beq\label{e9}
\left[\rho^2 \frac{\partial^2}{\partial \rho^2}+\rho
\frac{\partial}{\partial \rho}+ a^2 \rho^2 - m^2
\right]\varphi_A^\tau(r)=0 \eeq
which has as one of its solutions  the Bessel function of the
first kind that is regular at the origin
\begin{equation}\label{e10}
    \varphi^\tau_{A}(\rho)=C_{>}^\tau J_m(a\rho)
\end{equation}
 with $a=\varepsilon-u$ and $C_{>}^\tau$ is {the} constant of  normalization.
 The second
component of eigenspinor can be derived from
\eqref{e8} as
\begin{equation}\label{e11}
    \varphi^\tau_{B}(\rho)=- i C_{>}^\tau e^{-i\tau\varphi} J_{m-\tau}(a\rho).
\end{equation}
Finally in region $r>R$, the eigenspinors have the form
\begin{equation}\label{e12}
\Psi_{>}^\tau(\rho,\varphi)=C_{>}^\tau e^{im\varphi}\left(%
\begin{array}{c}
   J_{m}(a\rho) \\
  -ie^{-i\tau\varphi}J_{m-\tau}(a\rho) \\
\end{array}%
\right).
\end{equation}

As far as
the second region $r<R$ is concerned, the magnetic field $B$
forces the momentum operator
$\pi^{+}$ and $\pi^{-}$ to take the forms
\beq\label{e14}
\pi^{\pm}=-i\hbar v_Fe^{\pm
i\tau\varphi}\left(\frac{\partial} {\partial r}\pm\frac{\tau
i}{r}\frac{\partial}{\partial \varphi}\mp i\tau\frac{eB
r}{2\hbar}\right)
\eeq
in the symmetric gauge $\vec A= \frac{Br}{2} \vec e_\varphi$.
Solving the equation  $H\Psi_{<}^\tau=E\Psi_{<}^\tau$ we get
\begin{eqnarray}\label{e16}
&&
  \left(\frac{\partial}{\partial \rho}+\frac{m\tau}{\rho}+\tau \beta \rho\right)\varphi^\tau_{A}(\rho)=-(\varepsilon-u)\varphi^\tau_{B}(\rho)\\
  &&
  \label{e16bis}
  \left(\frac{\partial}{\partial \rho}+\frac{m\tau-1}{\rho}-\tau \beta \rho\right)\varphi^\tau_{B}(\rho)=(\varepsilon-u)\varphi^\tau_{A}(\rho)
\end{eqnarray}
where $\beta=\frac{eB}{2\hbar}R^2$ is a dimensionless parameter. These can be combined
to derive a second order differential equation
 \beq \label{e17}
 \left[\rho^2 \frac{\partial^2}{\partial
\rho^2}+\rho \frac{\partial}{\partial \rho}+ k_{\pm}^2 \rho^2 -
m^2-2\beta(m-\tau)+(\varepsilon-u_1)^2\rho^2-\beta^2\rho^4
\right]\varphi_A^\tau(\rho)=0
\eeq
which can be solved by considering the following ansatz
 \beq\label{e18}
 \varphi_{A}(\rho)=\rho^{|m|}e^{-\frac{\rho^2
\beta}{2}}\chi(\rho^{2}) \eeq
  yielding the confluent hypergeometric ordinary
differential equation
\beq\label{e19}
\left[x \frac{\partial^2}{\partial x^2
}+(b-x)\frac{\partial}{\partial x}-a\right]\chi(x)=0
\eeq
where  we put $x=\beta r^{2}$ and set the parameters
\beq\label{e20}
 b=1+|m|, \qquad
a=-\frac{(\varepsilon-u_1)^2 }{4 \beta}+\frac{m-\tau+|m|+1}{2}.
\eeq
Consequently, the solution is
 \beq \label{e21}
\varphi_A(\rho)=\rho^{|m|}e^{-\frac{\beta \rho^2}{2}}C_{<}^\tau
\tilde{M}\left(a,b,\beta \rho^2\right) \eeq
 $C_{<}^\tau$ is {the}  constant of normalization and
 $\tilde{M}(a,b,\beta \rho^2)$ are the confluent hypergeometric functions \cite{Abramowitz},
 {the subscript $<$ means solution for $r<R$}.
The second component can be extracted from
\eqref{e16} as
\beq \label{e22}
\varphi_B(\rho)=i\frac{C_{<}^\tau \rho^{|m|}e^{-\frac{\beta
\rho^2}{2}}e^{-i\tau\varphi}}{\varepsilon-u}\left[\left(\frac{\tau
m}{\rho}+\beta\tau \rho\right)
\tilde{M}\left(a,b,\beta\rho^2\right)-
a\tilde{M}\left(a+1,b+1,\beta \rho^2\right)\right]. \eeq
Combining all, we end up with the eigenspinors in the magnetic region $r<R$
\begin{equation}\label{e23}
\Psi_{<}^\tau(\rho,\varphi)=C_{<}^\tau \rho^{|m|}e^{-\frac{\beta
\rho^2}{2}}e^{im\varphi}\left(%
\begin{array}{c}
\tilde{M}\left(a,b,\beta \rho^2\right) \\
  \frac{ie^{-i\tau\varphi} }{\varepsilon-u}\left[\left(\frac{\tau
m}{\rho}+\beta\tau \rho\right)\tilde{M}(a,b,\beta
\rho^2)-a\tilde{M}\left(a+1,b+1,\beta \rho^2\right)\right]
\end{array}%
\right)
\end{equation}

Now we look for the energy levels of our system that
cannot be obtained {by} directly  solving the eigenvalue equation.
Nevertheless, we can still
apply the boundary condition
at the interface $\rho=1$ or $r=R$, namely
%
$\Psi_{>}^\tau(1)=\Psi_{<}^\tau(1)$. This operation yields
\beqar\label{e24}
  &&C_{>}^\tau J_{m}(a)=C_{<}^\tau e^{-\frac{\beta}{2}}\tilde{M}(a,b,\beta) \\
  &&C_{>}^\tau J_{m-\tau}(a)=-C_{<}^\tau \frac{e^{-\frac{\beta
}{2}}}{\varepsilon-u}\left[\left(\tau
m+\beta\tau\right)\tilde{M}(a,b,\beta)
-a\tilde{M}(a+1,b+1,\beta )\right]
\eeqar
which can be cast in matrix form as
\begin{equation}\label{e25}
M^\tau\left(
        \begin{array}{c}
          C_{>}^\tau \\
          C_{<}^\tau \\
        \end{array}
      \right)=0
      %
       \end{equation}
       such that the matrix is given by 
\beqar\label{e26}
M^\tau=
\left(
  \begin{array}{cc}
  J_{m}(a) & -e^{-\frac{\beta}{2}}\tilde{M}(a,b,\beta) \\
  J_{m-\tau}(a) &
  \frac{e^{-\frac{\beta}{2}}}{\varepsilon-u}\left[\left(\tau m+\beta\tau\right)\tilde{M}(a,b,\beta)-a\tilde{M}(a+1,b+1,\beta )\right]
  \end{array}
  \right).
\eeqar
The only allowed  energy levels are given by
$\det M^\tau=0$, which leads to the following eigenvalue equation
\beq
J_{m}(a)\ \frac{e^{-\frac{\beta}{2}}}{\varepsilon-u}\left[\left(\tau m+\beta\tau\right)\tilde{M}(a,b,\beta)-a\tilde{M}(a+1,b+1,\beta )\right]
+ J_{m-\tau}(a) \ e^{-\frac{\beta}{2}}\tilde{M}(a,b,\beta).
\eeq
Recall that the energy levels are embedded in the parameter $a=\varepsilon-u$.
This equation
will
be numerically solved
to extract the energy levels,
which will then allow us to study the basic properties of our system. In fact, 
we will discuss such levels under various configurations
of the physical parameters such as
the  quantum dot radius $R$,
magnetic field $B$ and electrostatic potential.

\section{Numerical results}\label{sec:results}

In Figure \ref{f2}, we show
the energy levels  as a function of the quantum dot radius $R$ for $b=10$ T and
three different values of angular quantum number such that
(a): $m=-1$, (b): $m=0$, (c): $m=1$  with  $U=0$ meV
and (d): $m=-1$, (e): $m=0$, (f): $m=1$  with  $U=100$ meV. Note that, full  and dashed lines are describing energy associated with
 the two valleys $K$ ($\tau=1$) and $K'$ ($\tau=-1$), respectively. We observe that when $R\longrightarrow 0$, the energies of  $K$ ($\tau=1$) and $K' $($\tau=-1$) are degenerate for all $\tau$, which means that $E(m,\tau)=E(m,-\tau$). However when $R$ increases, two sets of energy levels appear, one shows the
symmetry $E(m,\tau)=E(m,-\tau)$
and approaches  the Landau levels (LLs) corresponding to
graphene QD \cite{Mirzakhani16}, the other one shows a lack of
 symmetry $E(m,\tau)\neq E(m,-\tau)$. It is clearly seen that an energy gap is opened for a non zero angular momentum $m$
as shown in Figures \ref{f2}(a,c,d,f) and there is a zero energy  for $m=0 $ as in  Figures \ref{f2}(b,e). On the other hand,  the influence of potential $U$ makes it possible to move the energy levels vertically by $U$, namely
we have
\beq\lb{e288}
{E(m,\tau,U)=E(m,\tau)+U.}
\eeq

 The energy levels as a function of
 the magnetic field {are} presented in Figure \ref{f3} for 
 $R=70$ nm,
 (a): $m=-1$, (b): $m=0$,   (c): $m=1$ with $U=0$ meV
 and (d): $m=-1$, (e): $m=0$, (f): $m=1$  with $U=100$ meV.
   For small magnetic field ($B\longrightarrow 0$), we observe that the energy levels display a continuum energy band
 and also there are many degenerate zero-energy states corresponding to all angular momenta $m$ for both $K$ and $K'$, i.e. $E(m,\tau)=E(m,-\tau)$ \cite{Mirzakhani16}. Now by increasing 
 $B$, one notices that the degeneracy of the energy levels is lifted
 and then we have $E(m,\tau)\neq E(m,-\tau)$, which  finally  connects to the LLs of graphene subject to magnetic field  \cite{Schnez08} 
  \begin{equation}\label{e28}
E_{nm}-U=\pm 
{E_0}\sqrt{2\beta(2n+m+|m|+1-\tau)}, \qquad n=0, 1,2, \cdots
\end{equation}
which can be derived from the relation between the Laguerre  function and {confluent hypergeometric function $\tilde M$}
under suitable conditions, more detail can be found in
\cite{Recher009}.
  There are some other features, e.g. 
  for $B\neq 0$ an energy gap appears between the conduction and valence bands, which is  quantum number $m$ dependent. The behavior of the potential on the electronic properties of the energy levels for  $K$ and $K'$ shows that 
  \eqref{e288}
  is qualitatively similar to the behavior seen in
  Figure \ref{f2}.\\

\begin{figure}[h]
  \center
  \includegraphics[width=5.4cm]{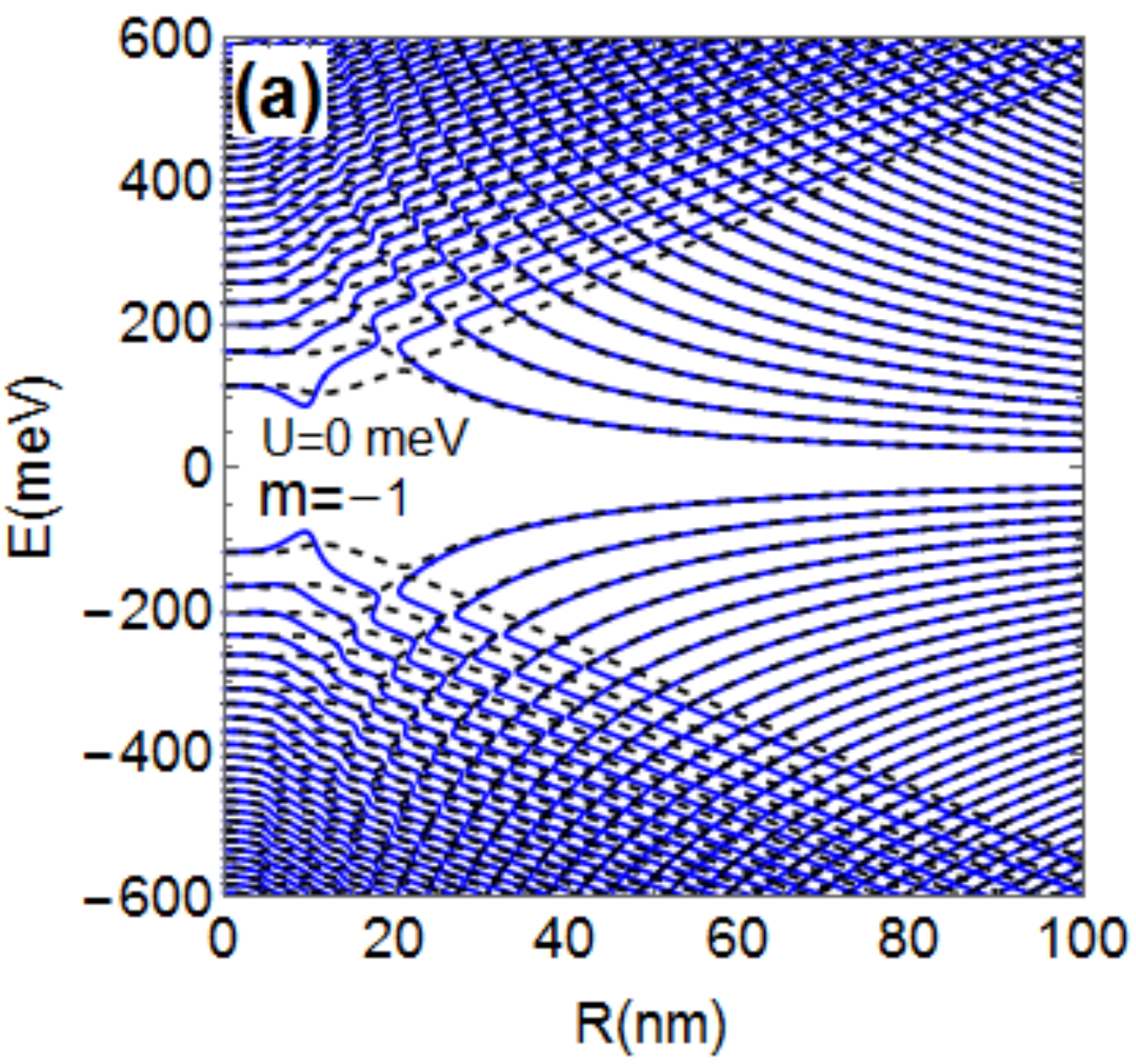}
  \includegraphics[width=5.5cm]{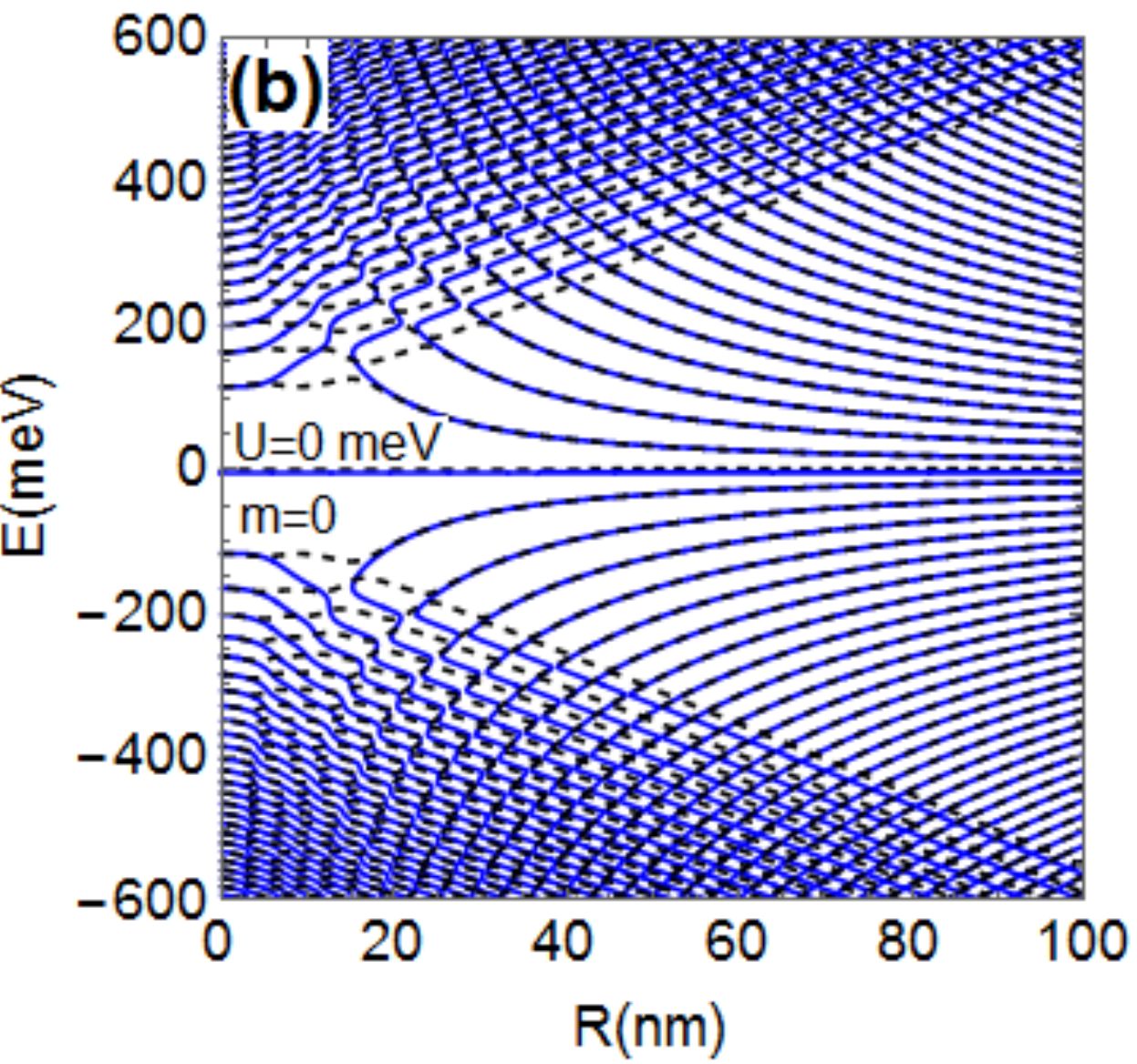}
  \includegraphics[width=5.4cm]{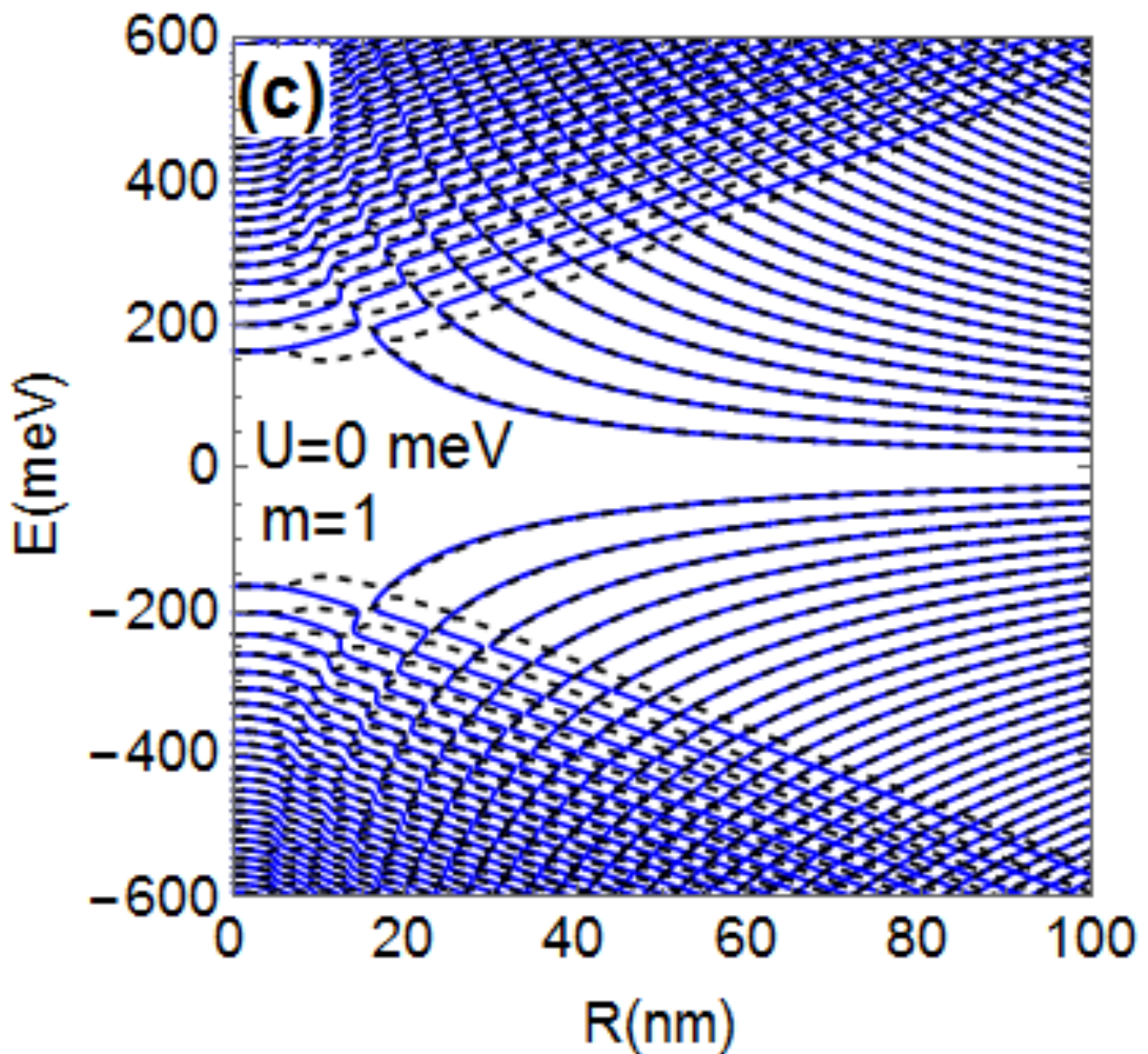}
  \includegraphics[width=5.5cm]{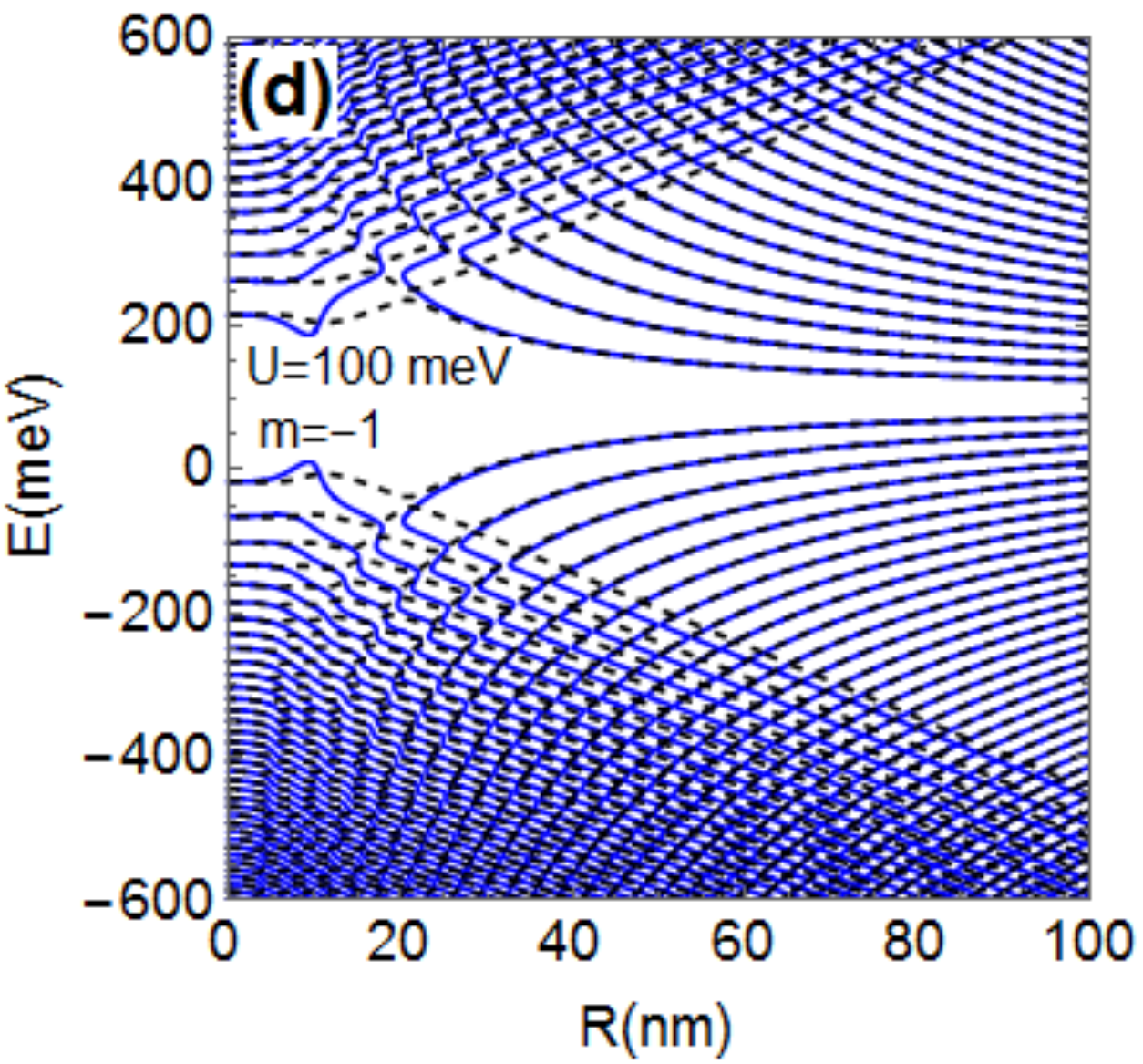}
  \includegraphics[width=5.5cm]{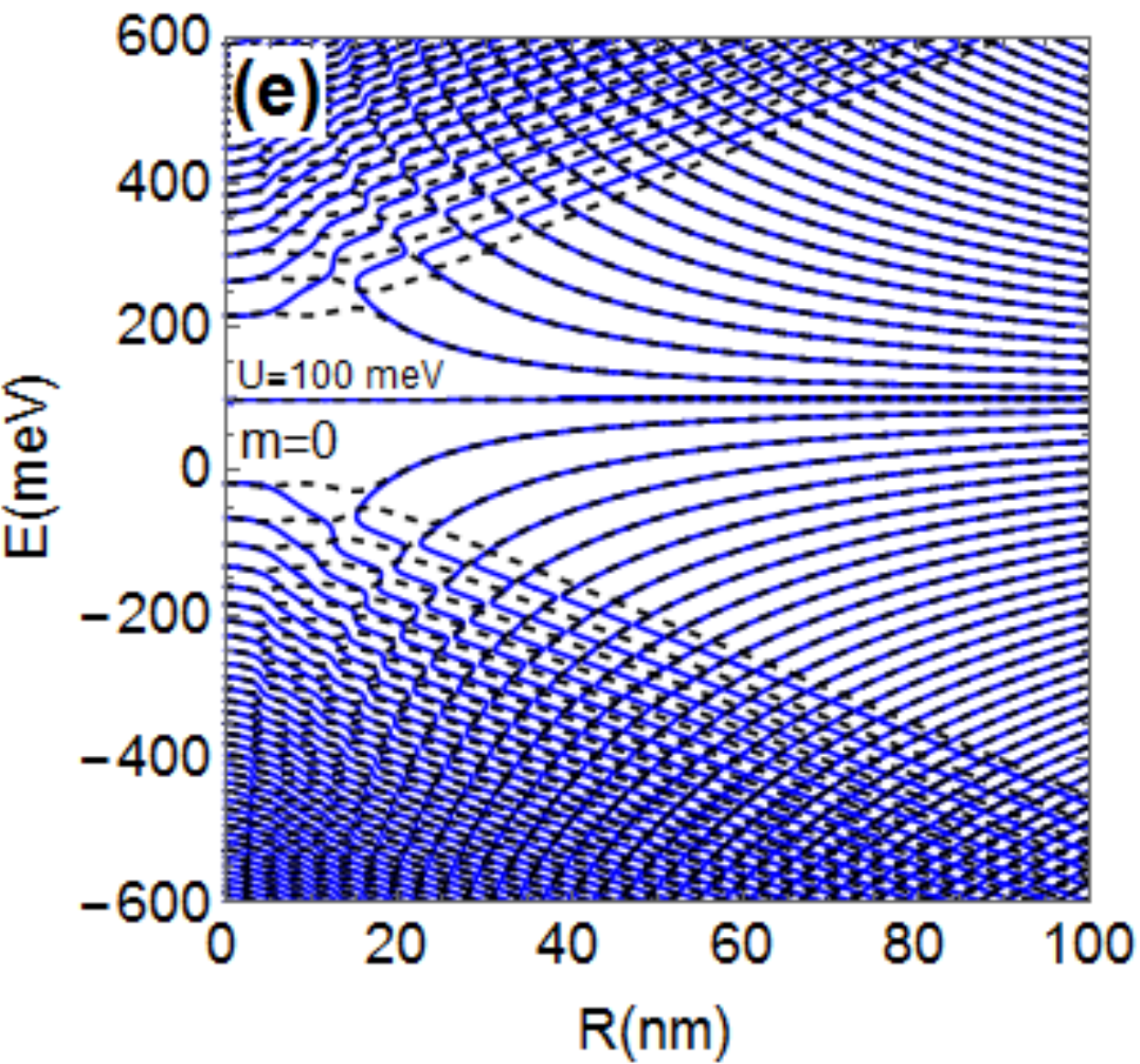}
  \includegraphics[width=5.5cm]{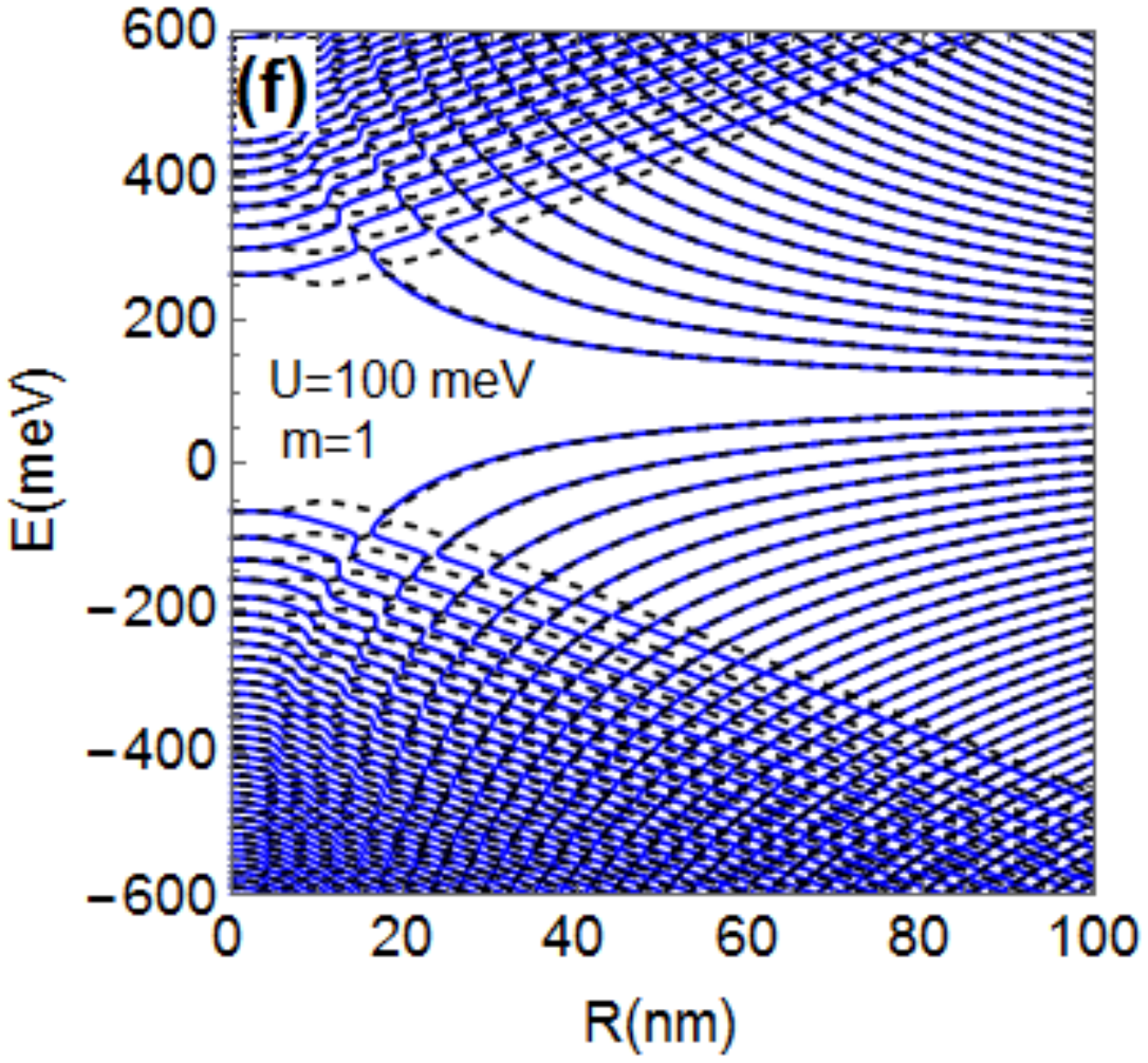}
  \caption{\sf(color online) Energy levels as a function of the quantum dot radius $R$ for $B=10$ T. (a): $m=-1$, (b): $m=0$, (c): $m=1$ with $U=0$ meV. (d): $m=-1$, (e): $m=0$, (f): $m=1$ with $U=100$ meV.  $\tau=1$ for blue line and $\tau=-1$ for red dashed line.
  \label{f2}}
\end{figure}
\begin{figure}[H]
  \center
  \includegraphics[width=5.6cm]{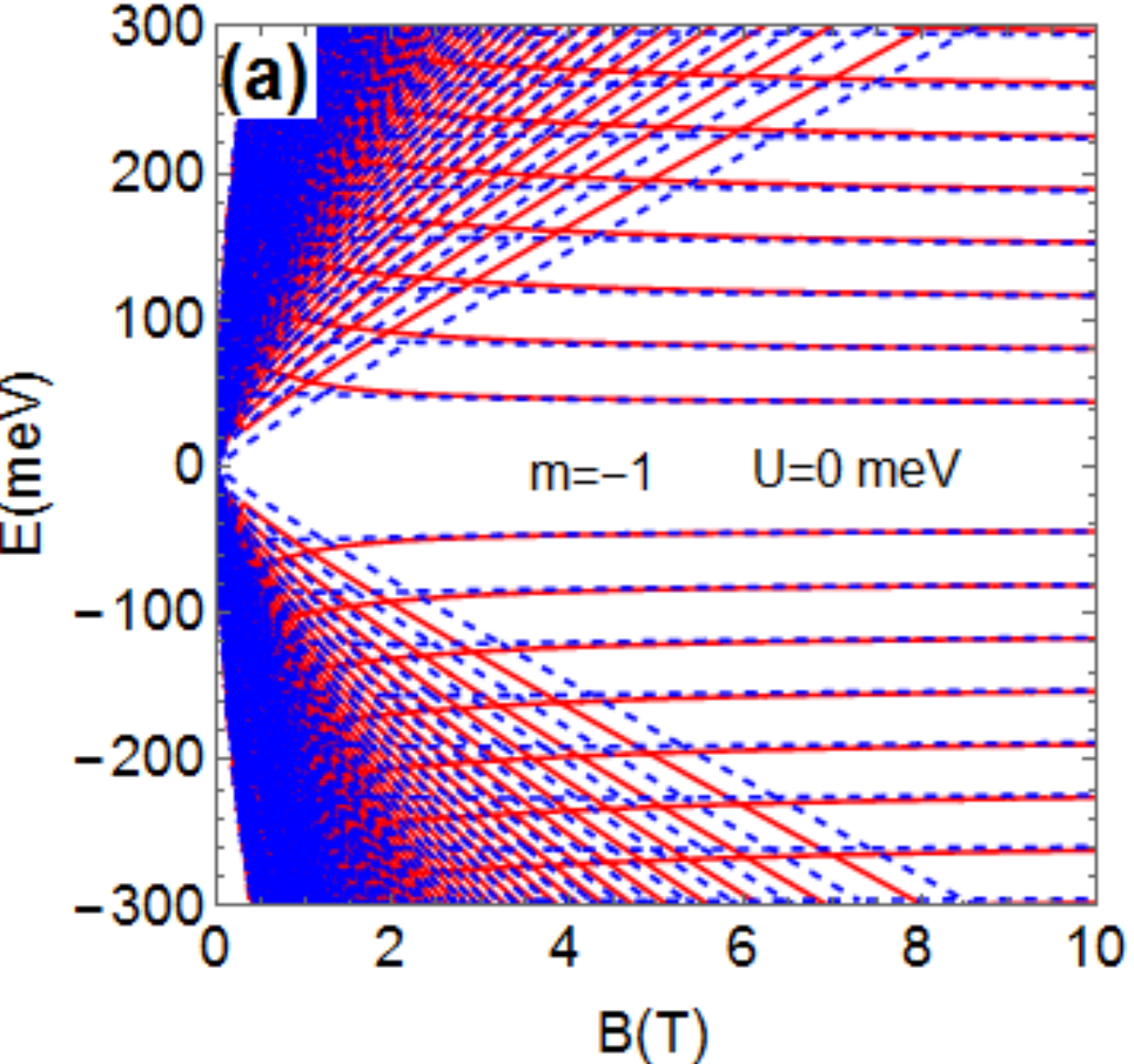}\includegraphics[width=5.6cm]{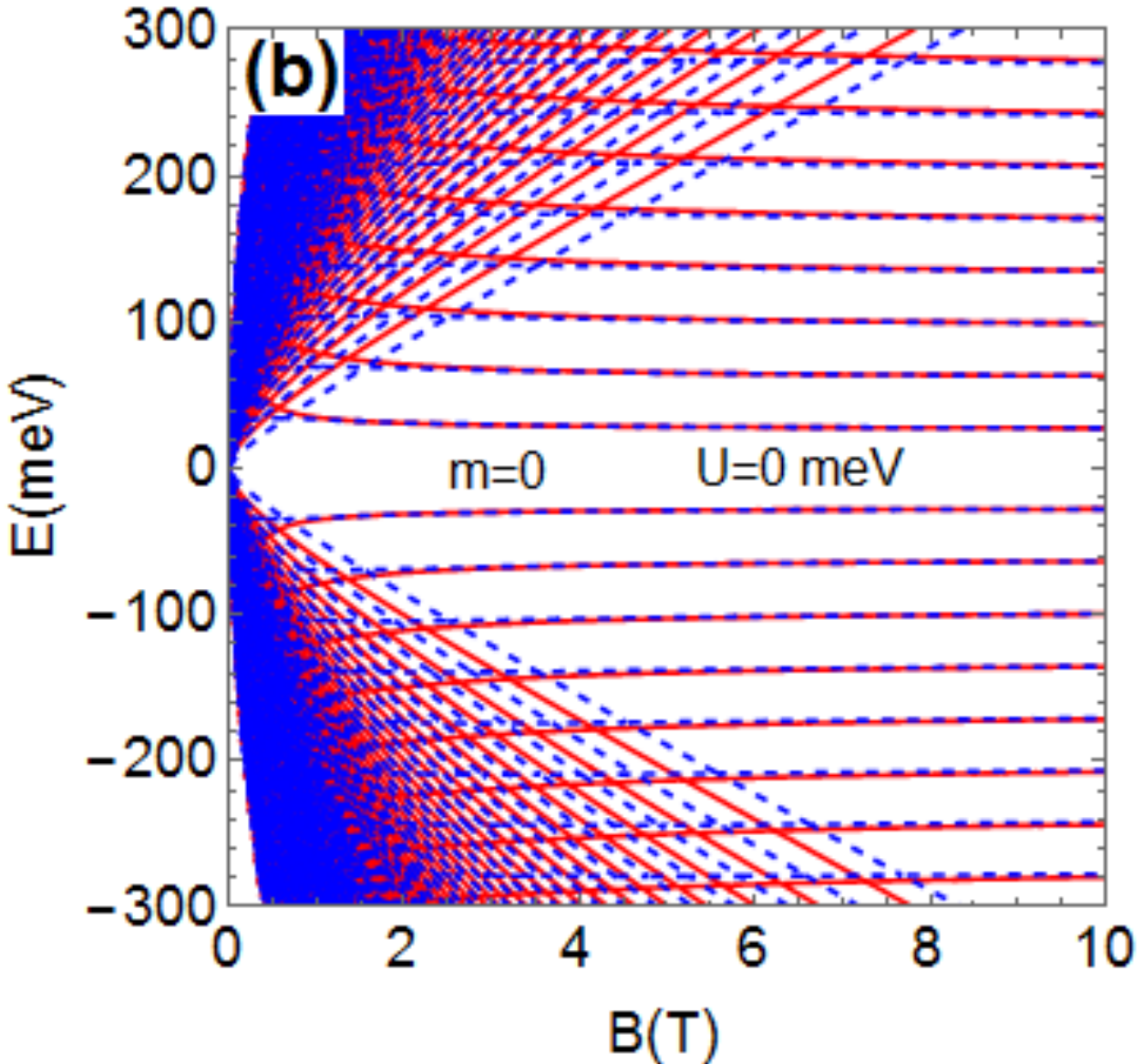}\includegraphics[width=5.6cm]{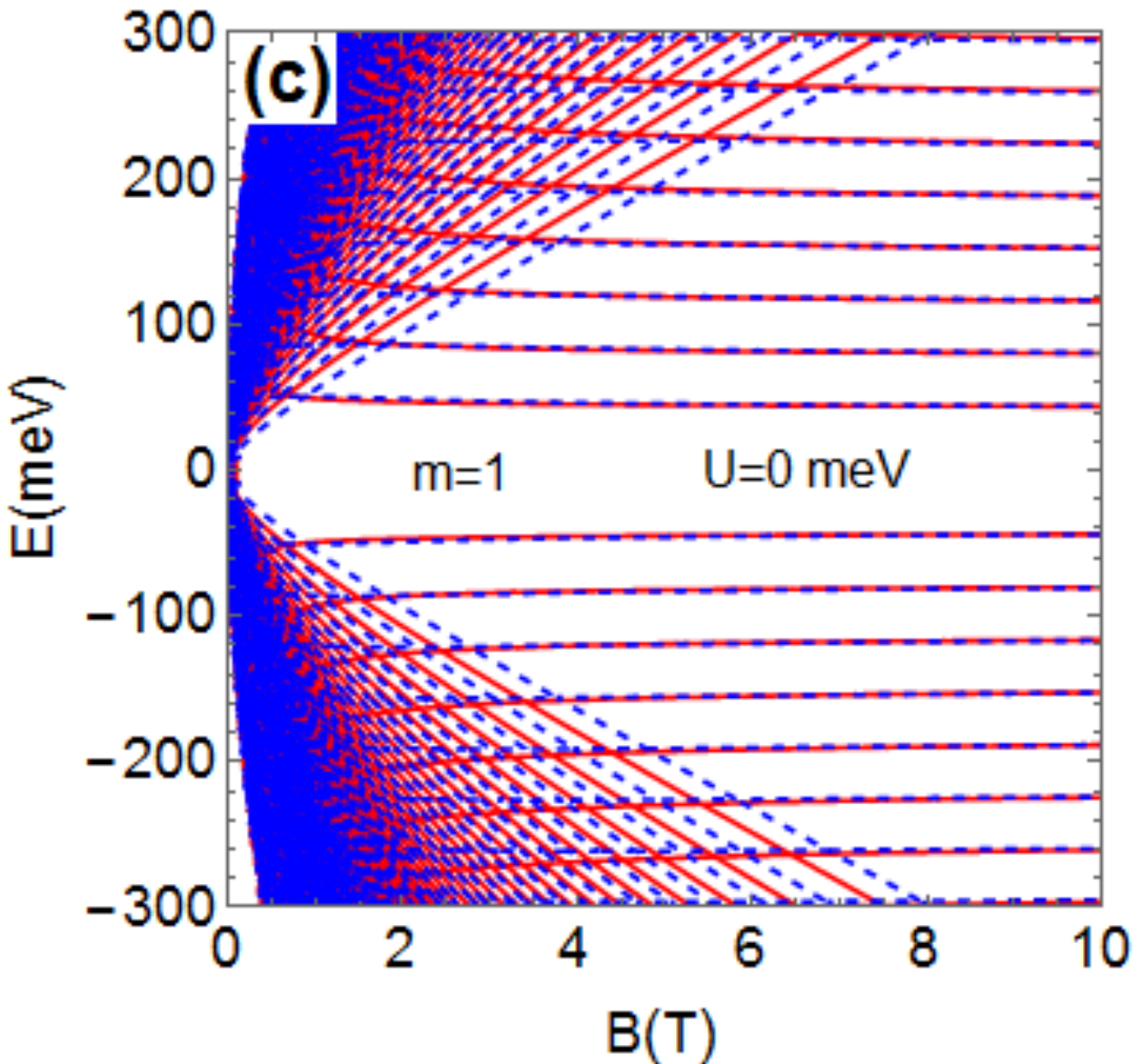}
  \includegraphics[width=5.6cm]{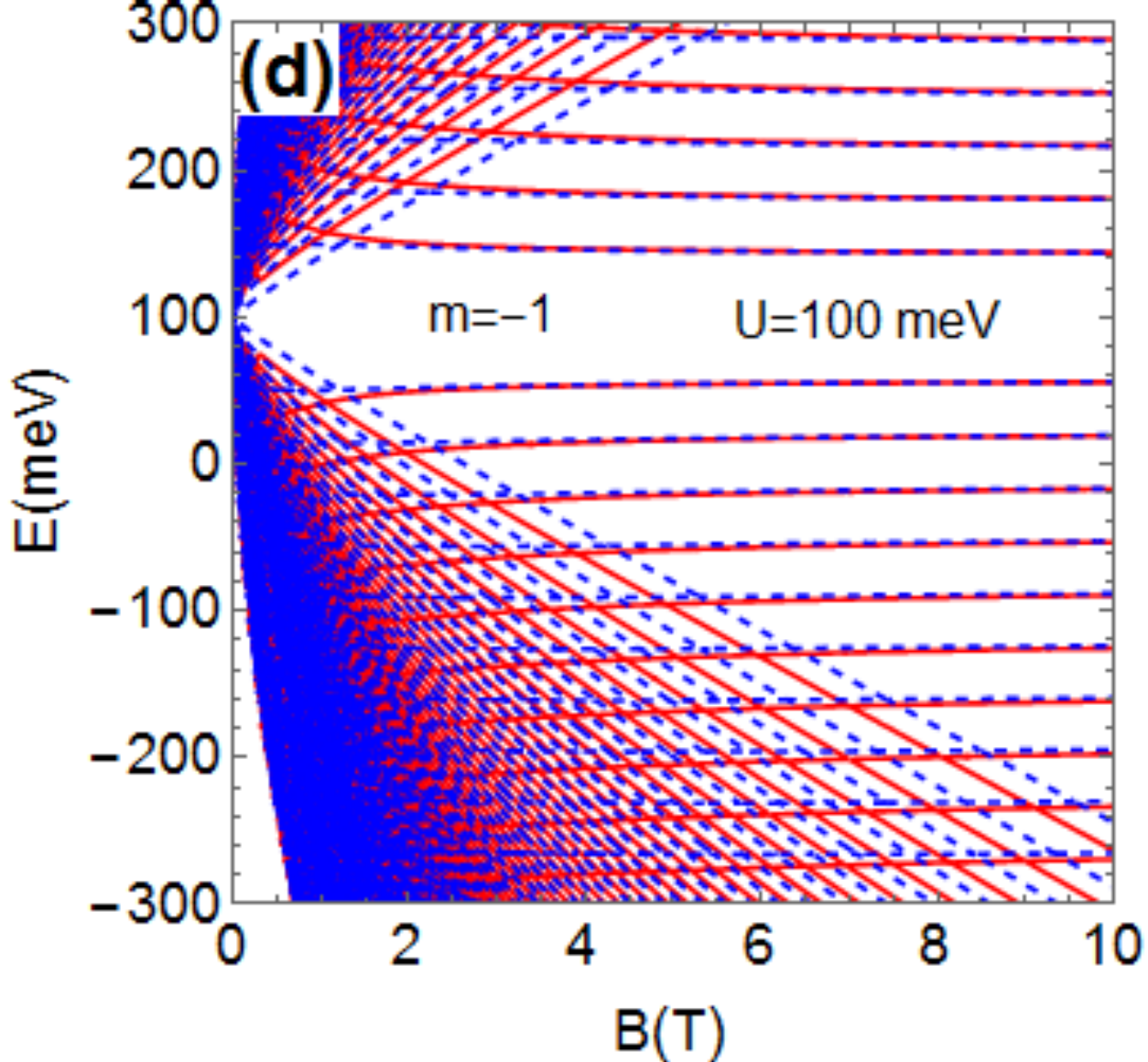}\includegraphics[width=5.6cm]{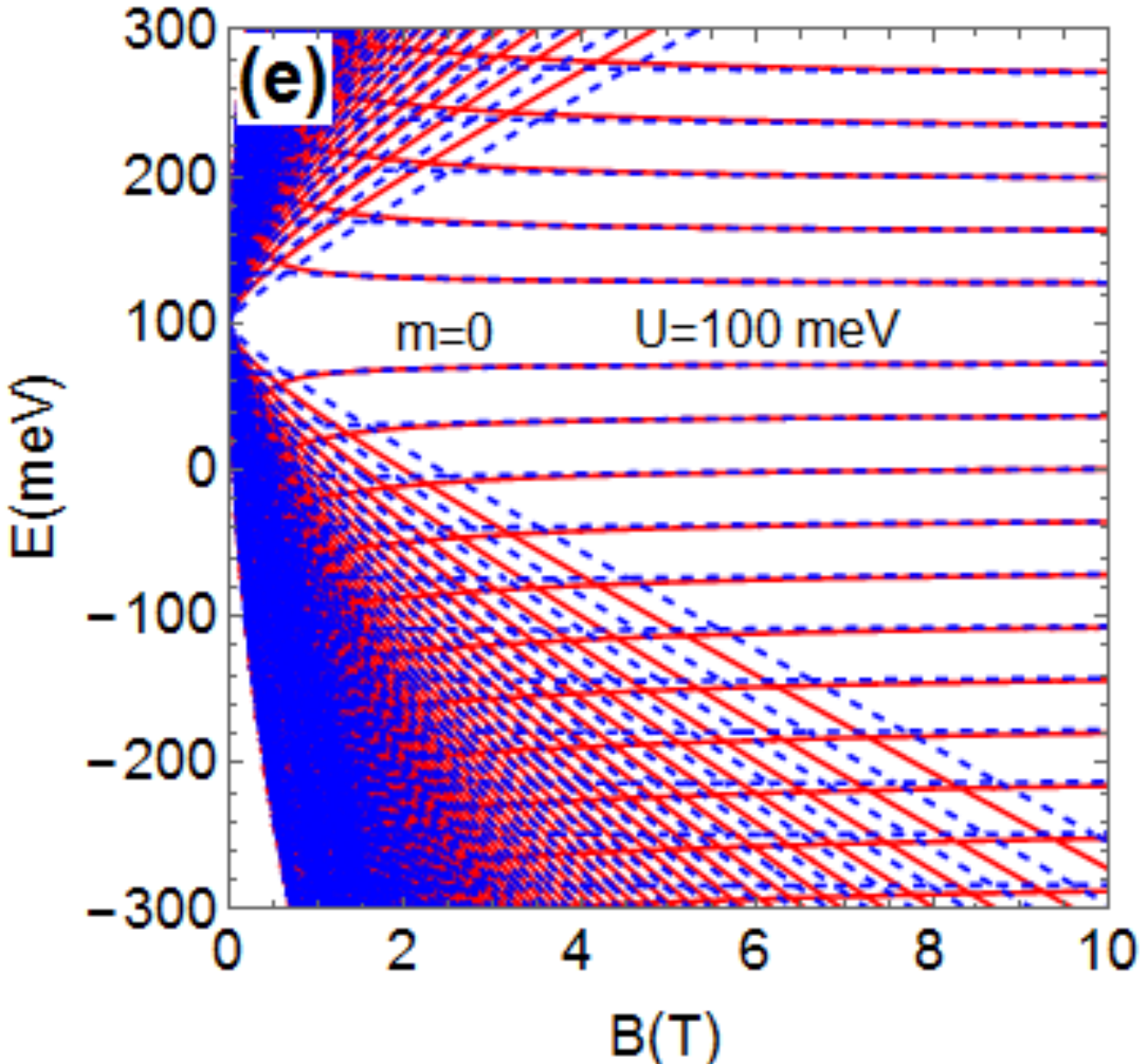}\includegraphics[width=5.6cm]{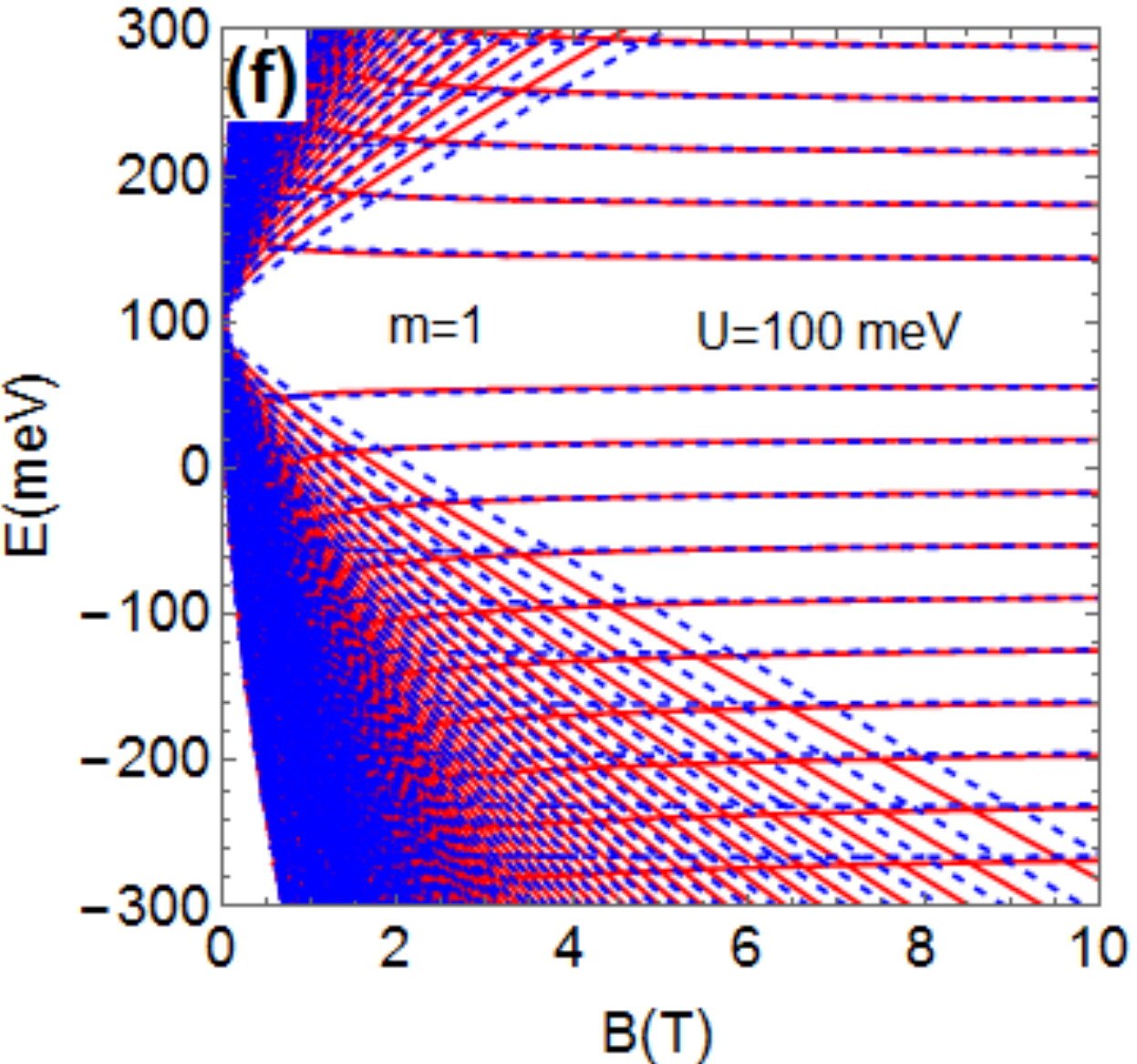}
  \caption{\sf (color online) Energy levels as a function of the magnetic field $B$ with $R=70$ nm. (a): $m=-1$, (b): $m=0$, (c): $m=1$ for $U=0$ meV. (d): $m=-1$, (e): $m=0$, (f):   $m=1$ for $U=100$ meV. Blue color for $\tau=1$ and red dashed color for $\tau=-1$. \label{f3}}
\end{figure}
\begin{figure}[h!]
  \center
  \includegraphics[width=8.2cm]{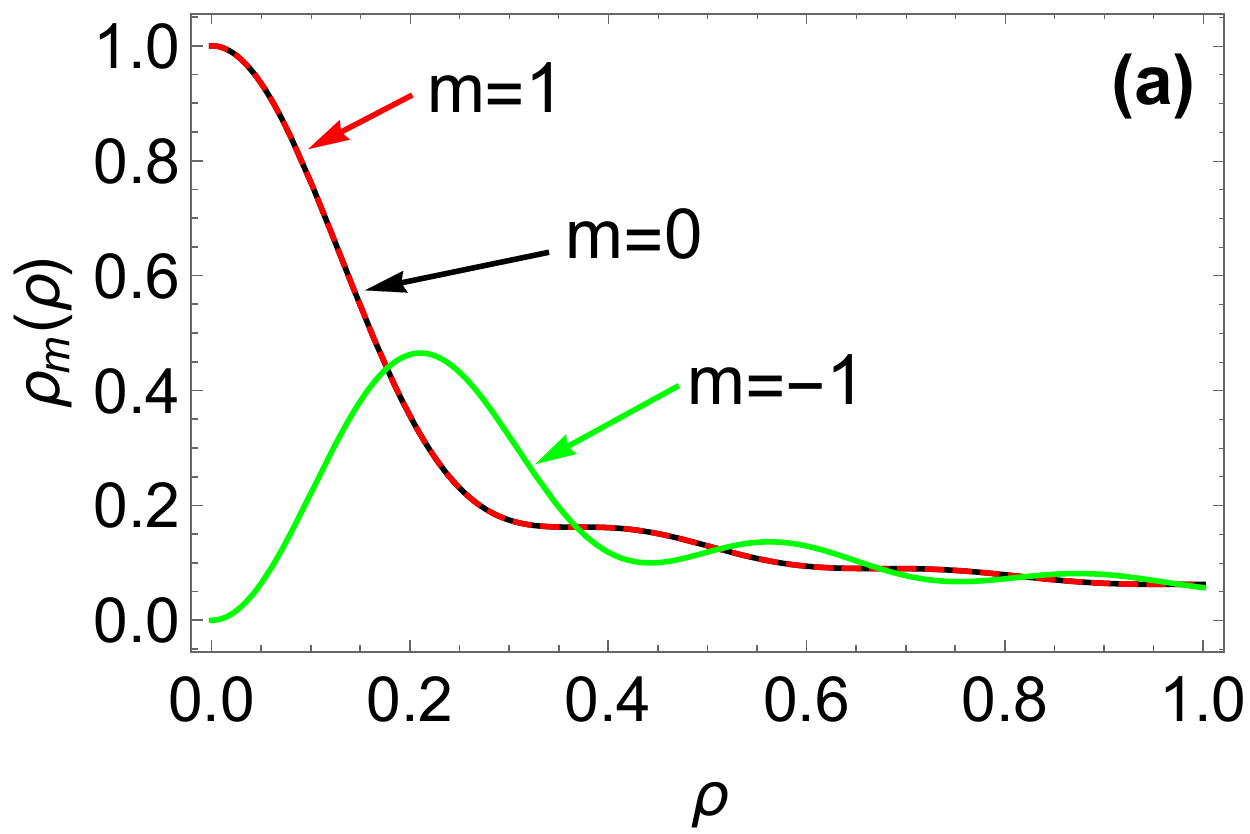}\includegraphics[width=8.2cm]{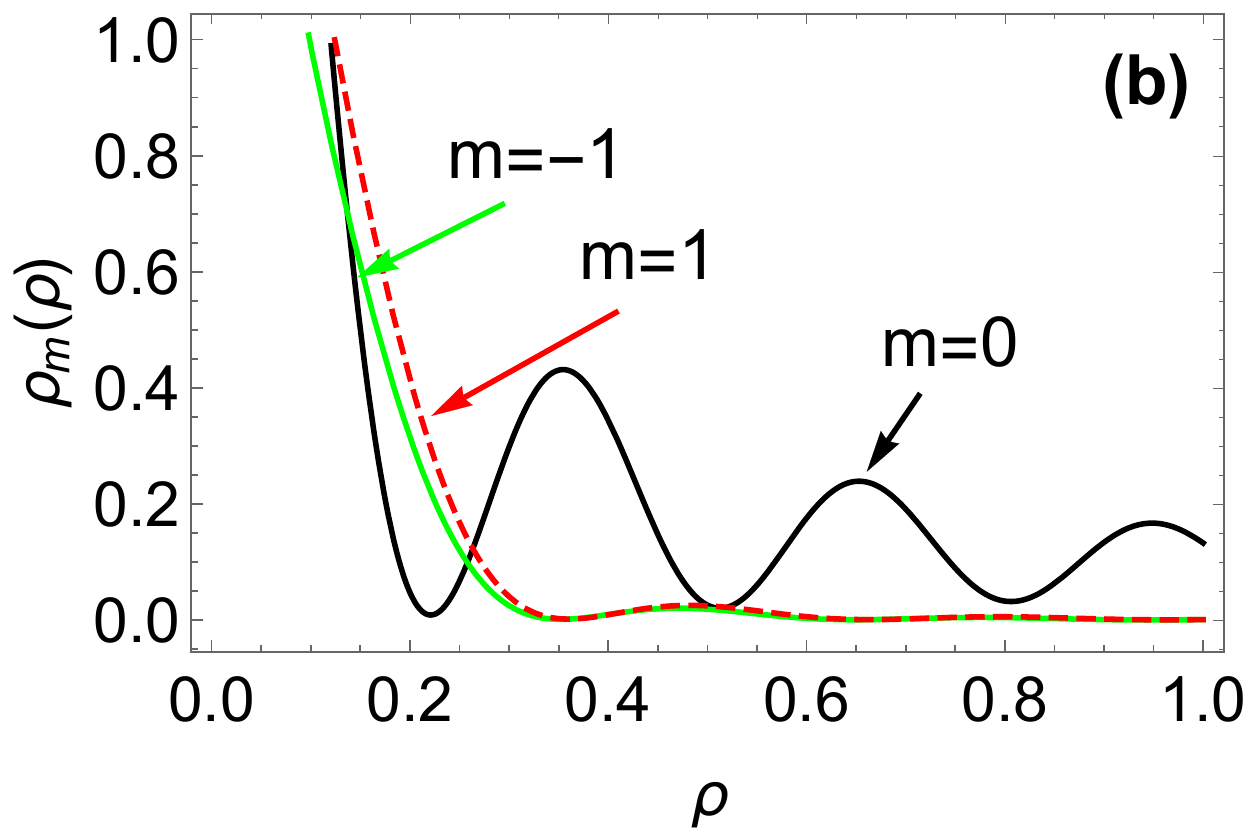}
  \caption{\sf (color online)  Radial probability {$\rho_{m}(\rho)$ as a function of the quantum dot radius $\rho=\frac{r}{R}$} with $U=100$ meV. (a): $B=0$ T, (b): $B=0.5$ T, for $\tau=\pm 1$ and $E=0.1$ {meV}. Total angular quantum number $m=-1$ (green), $m=0$ (back) and $m=1$ 
  (red). \label{f4p}}
\end{figure}

The radial probability {$\rho_{m}(\rho)$} as a function of the QD radius {$\rho$} are shown in Figure
\ref{f4p}, for $U=100$ meV, with $E=0.1$ {meV}, total angular quantum number $m=-1$ (green), $m=0$ (back) and $m=1$ (red). In the absence of the magnetic field (Figure \ref{f4p}(a)), for 
$m=0$ (black) and $m=1$ (red), we  clearly see that the maxima of the radial probability {$\rho_{m}(\rho)$} close to {$\rho=0$}. This maximum corresponds to the electron state strongly trapped in the quantum dot, but with the increase of {$\rho$}, {$\rho_{m}(\rho)$} tends to $0$. On the other hand, for $m=-1$, we have zero radial probability in the vicinity of {$\rho=0$}, however when {$\rho$} increases {$\rho_{m}(\rho)$} increases to a maximum value around the point {$\rho=0.2$} and after that it 
has a damped oscillatory behavior.
For a non-zero magnetic field Figure \ref{f4p}(b)), the radial probability {$\rho_{m}(\rho)$} starts its behavior with a maximum value at point from {$\rho=0.15$} for each value of  $m$. Note that for $m=0$ and $m=-1$, the probability decreases with the increase of {$\rho$},  in particular we have {$\rho_{m}(\rho)$} from {$\rho=0.35$}. For $m=0$, the radial probability has an approximately oscillatory dependence damped particularly for large {$\rho$}.

  In Figure \ref{f4}, we show the energy levels as {a} function of the confining potential $U$ for $R=70$ nm and $B=10$ T with (a): $m=-1$, (b): $m=0$, (c): $m=1$.
We observe that they have a linear form and are twofold degenerate due to the symmetry  $E(m,\tau)=E(-m,-\tau)$ for $m\neq 0$ and $E(m,\tau)=E(m,-\tau)$ for $m=0$. It is clearly seen that  an energy gap is opened for all {the values} of $m$ between conduction and valance bands.\\

\begin{figure}[h!]
  \center
  \includegraphics[width=5.6cm]{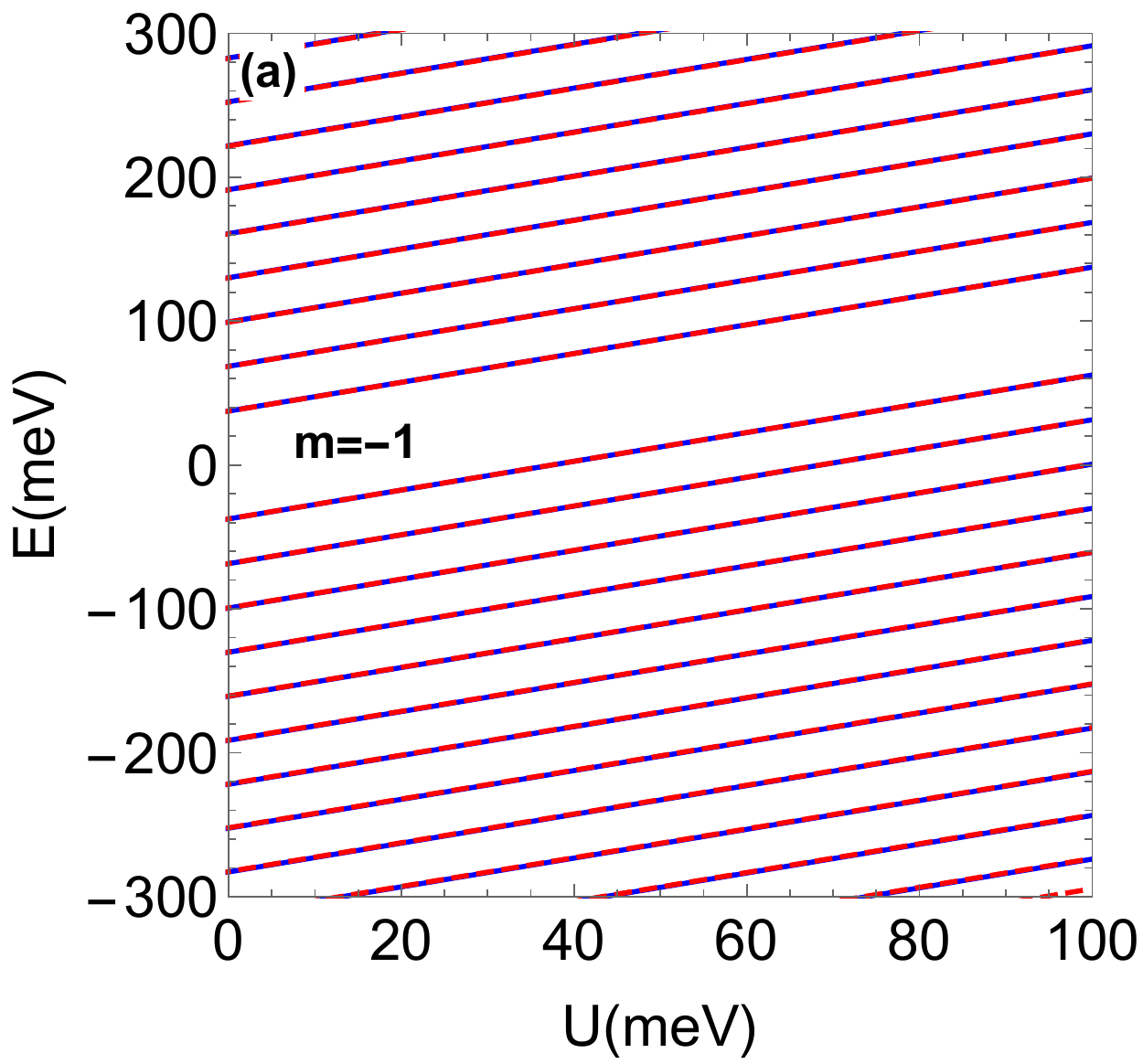}\includegraphics[width=5.6cm]{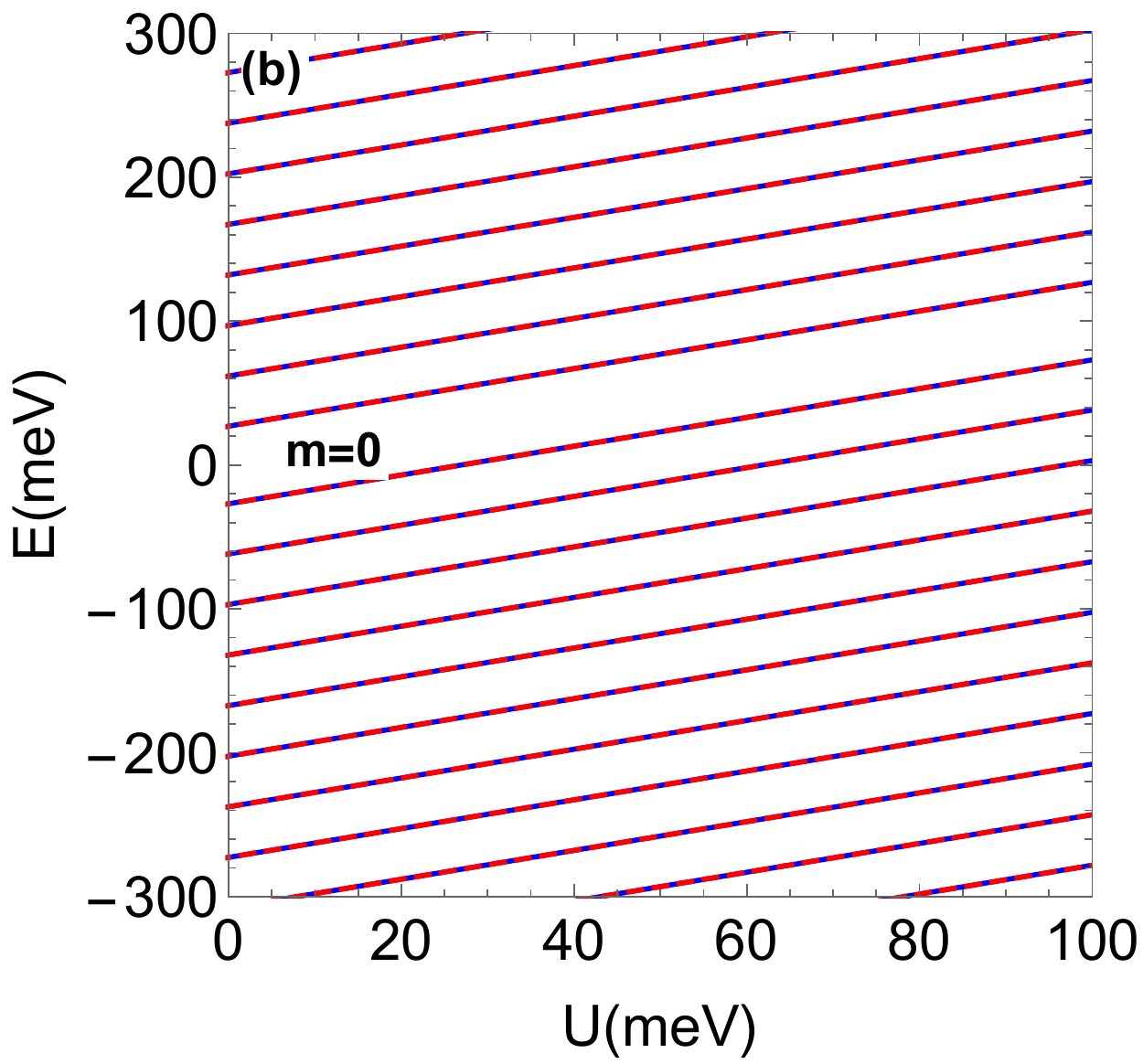}\includegraphics[width=5.6cm]{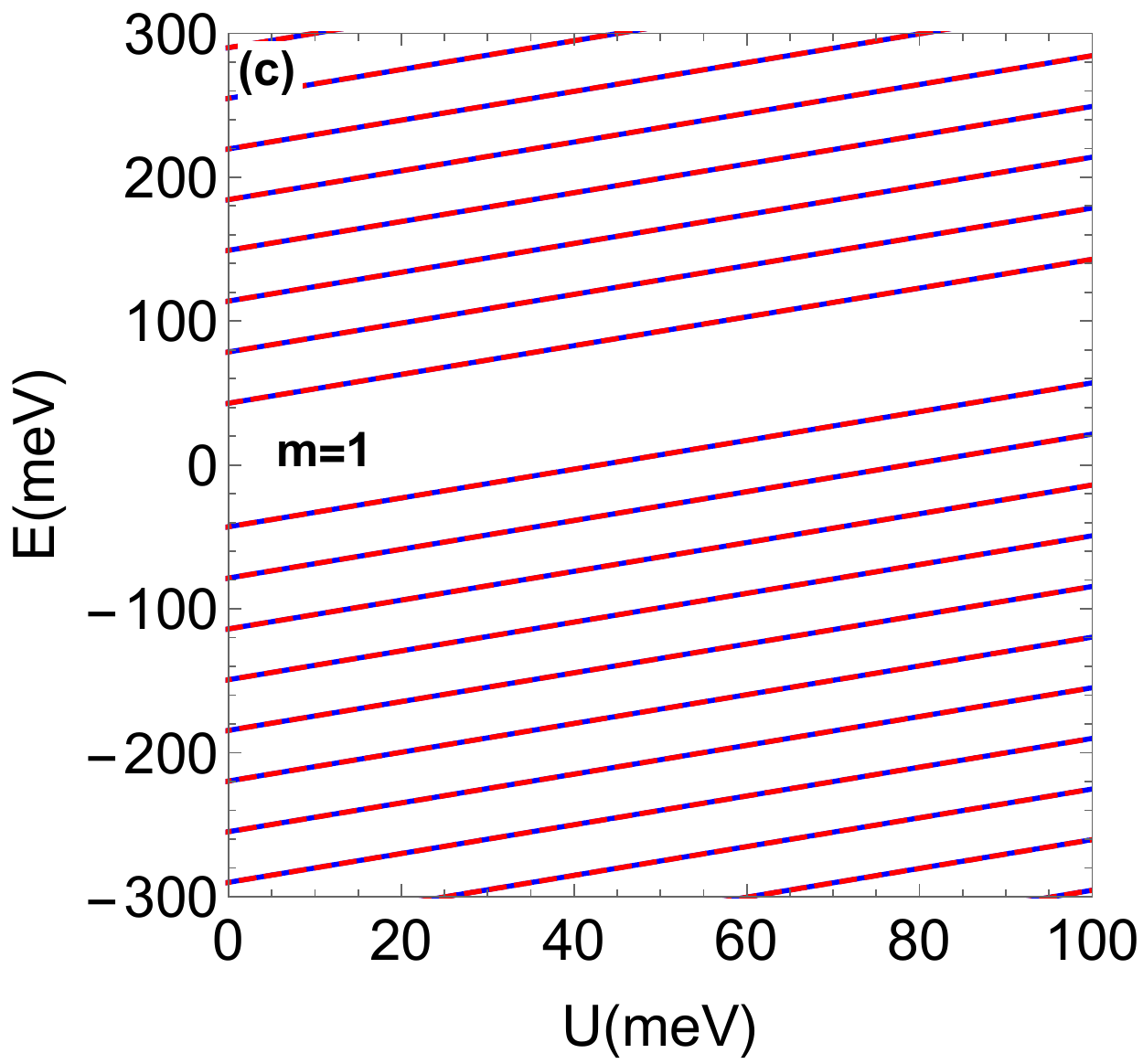}
  \caption{\sf (color online) Energy levels as {a} function of the potential $U$ with $B=10$ T. (a): $m=-1$, (b): $m=0$, (c): $m=1$  for $R=70$ nm. Blue color for $\tau=1$ and red dashed color for $\tau=-1$. \label{f4}}
\end{figure}

\begin{figure}[h!]
  \center
  \includegraphics[width=8.5cm]{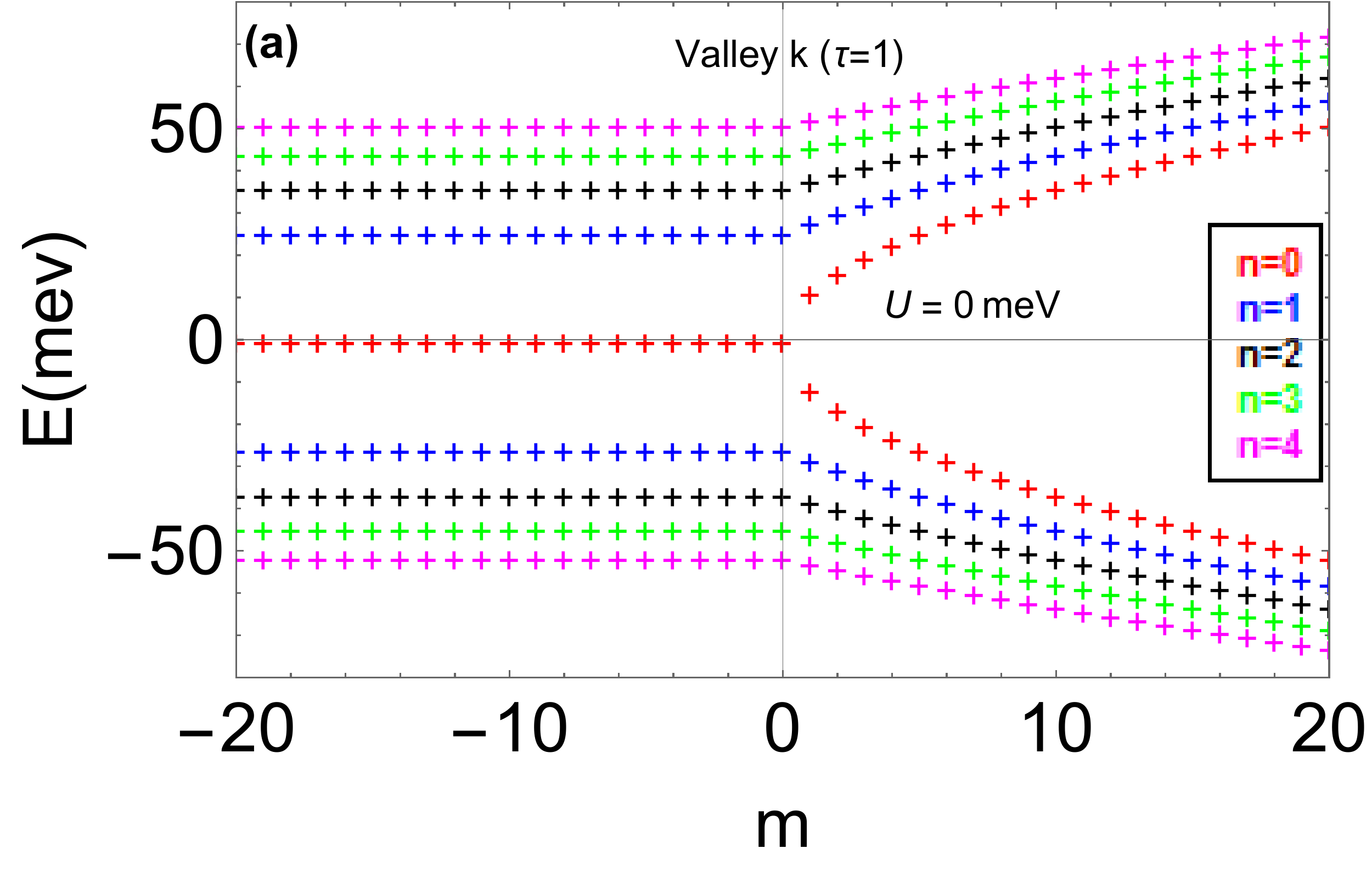}\includegraphics[width=8.5cm]{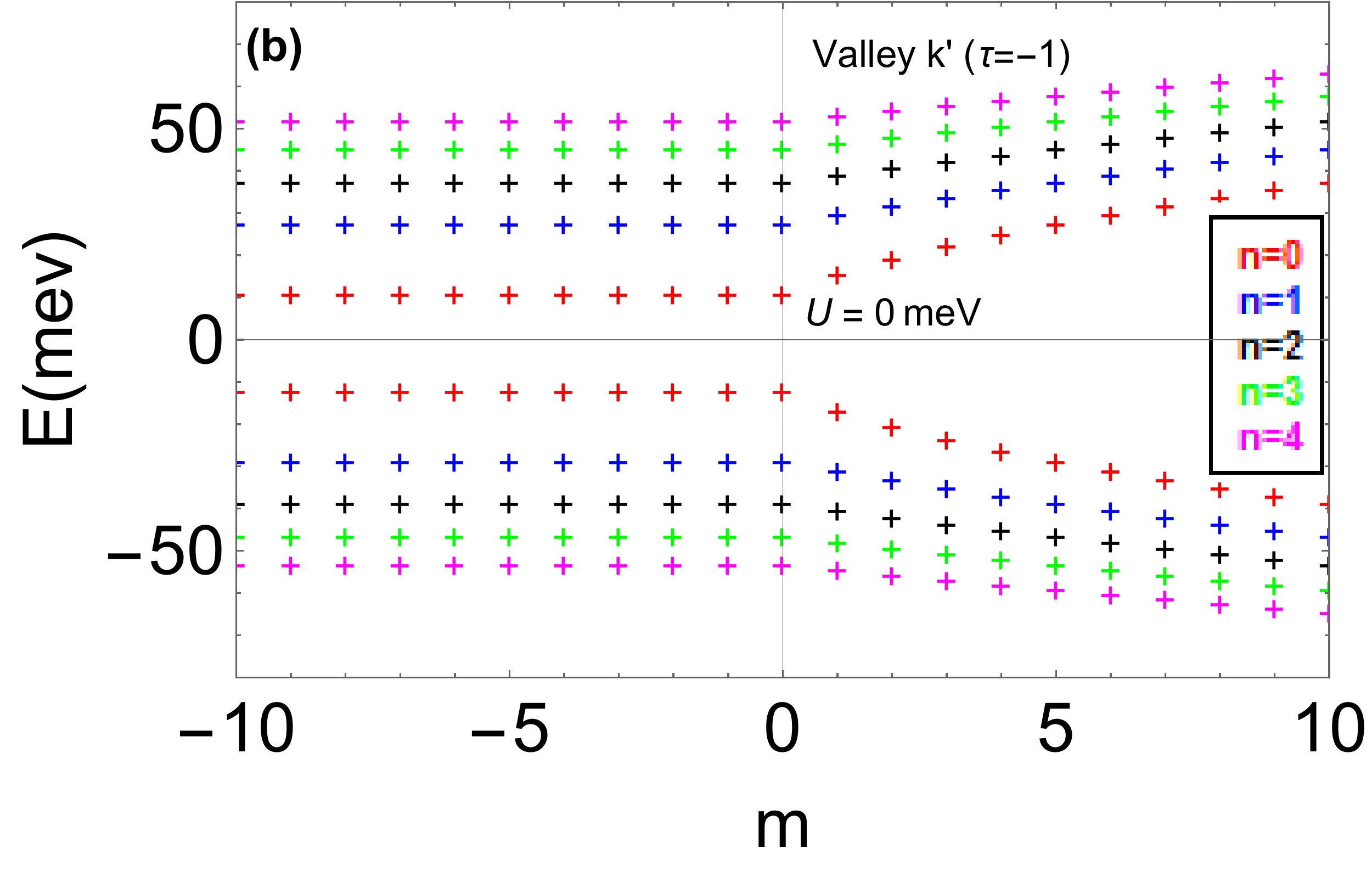}
  \includegraphics[width=8.5cm]{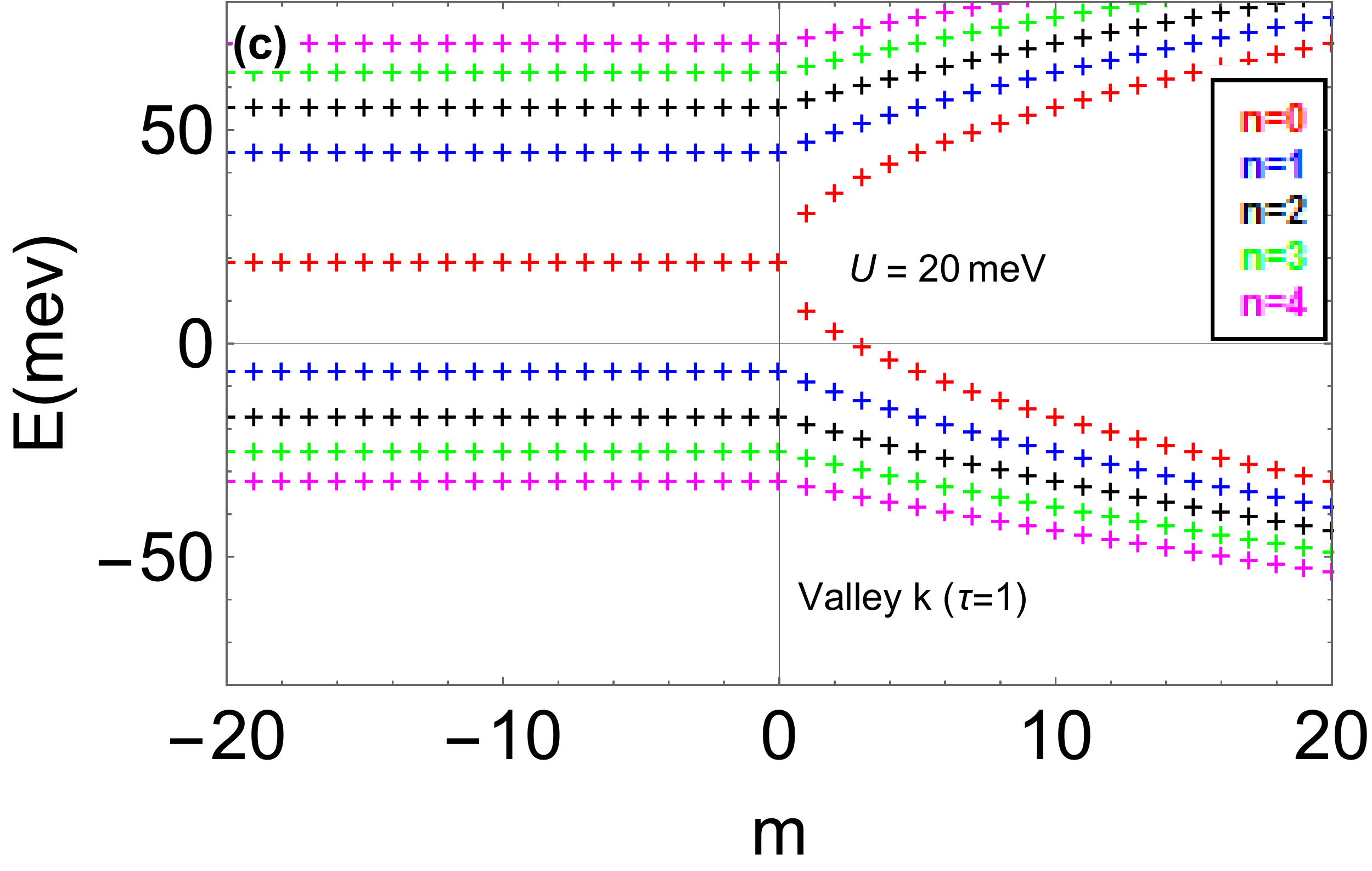}\includegraphics[width=8.5cm]{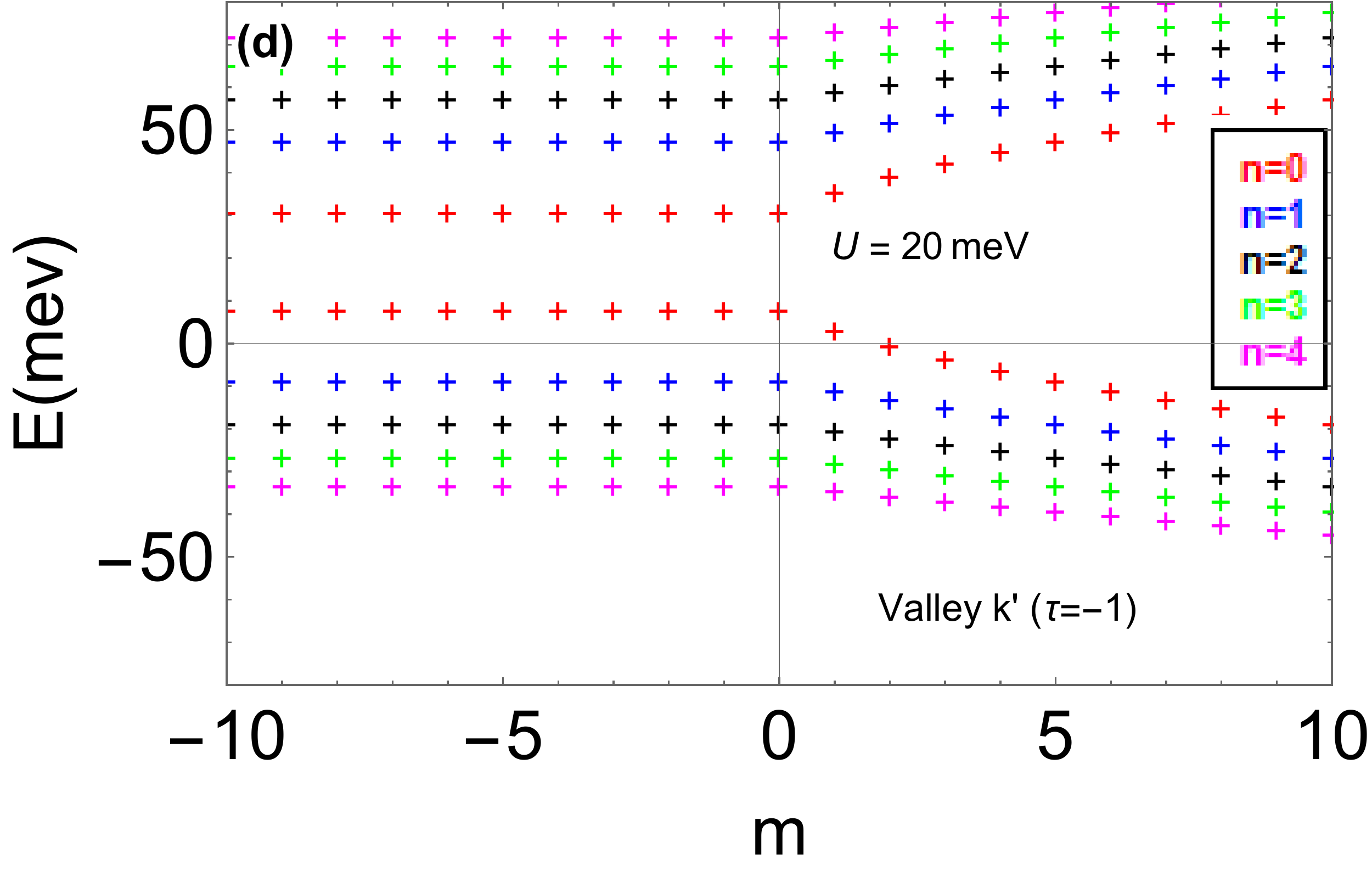}
  \caption{\sf (color online) {Energy levels} $E_{nm}$, given in \eqref{e28}, as
  {a} function of angular momentum $m$ with  $B=10$ T for Landau levels $n=0,\cdots, 4$. (a): $U=0$ meV, (c): $U=20$ meV for $\tau=-1$, (b): $U=0$ meV, (d): $U=20$ meV for $\tau=1$.\label{f5}}
\end{figure}

The resulting eigenenergies $E_{nm}$ \eqref{e28} are plotted in Figure \ref{f5} for $U=0$ in  {the} panels (a,b) and $U=20$ meV in  {the} panels (c,d)
with Landau levels $n=0,\cdots 4$.
It is noticed
that the two valleys are non degenerate in the eigenenergy spectra 
$E_{nm}(\tau)\neq E_{nm}(-\tau)$.
The eigenenergies exhibit asymmetric behavior with respect to the sign of the quantum number $m$ i.e. $E_{nm}(m,\tau)\neq E_{nm}(-m,\tau)$. For the valley $K$ the energy spectrum $E_{nm} (\tau=1)$ has a zero gap, but for the valley $K'$ the energy spectrum $E_{nm} (\tau=-1)$ has a non-zero gap. It is clearly seen that  the existence of the potential $U$ allows to move the energy levels vertically as shown in panels (c,d). For larger $|m|$, larger gap are observed between valance and conduction bands for the two valleys $\tau=\pm1$.\\

\begin{figure}[h!]
  \center
  \includegraphics[width=4.2cm]{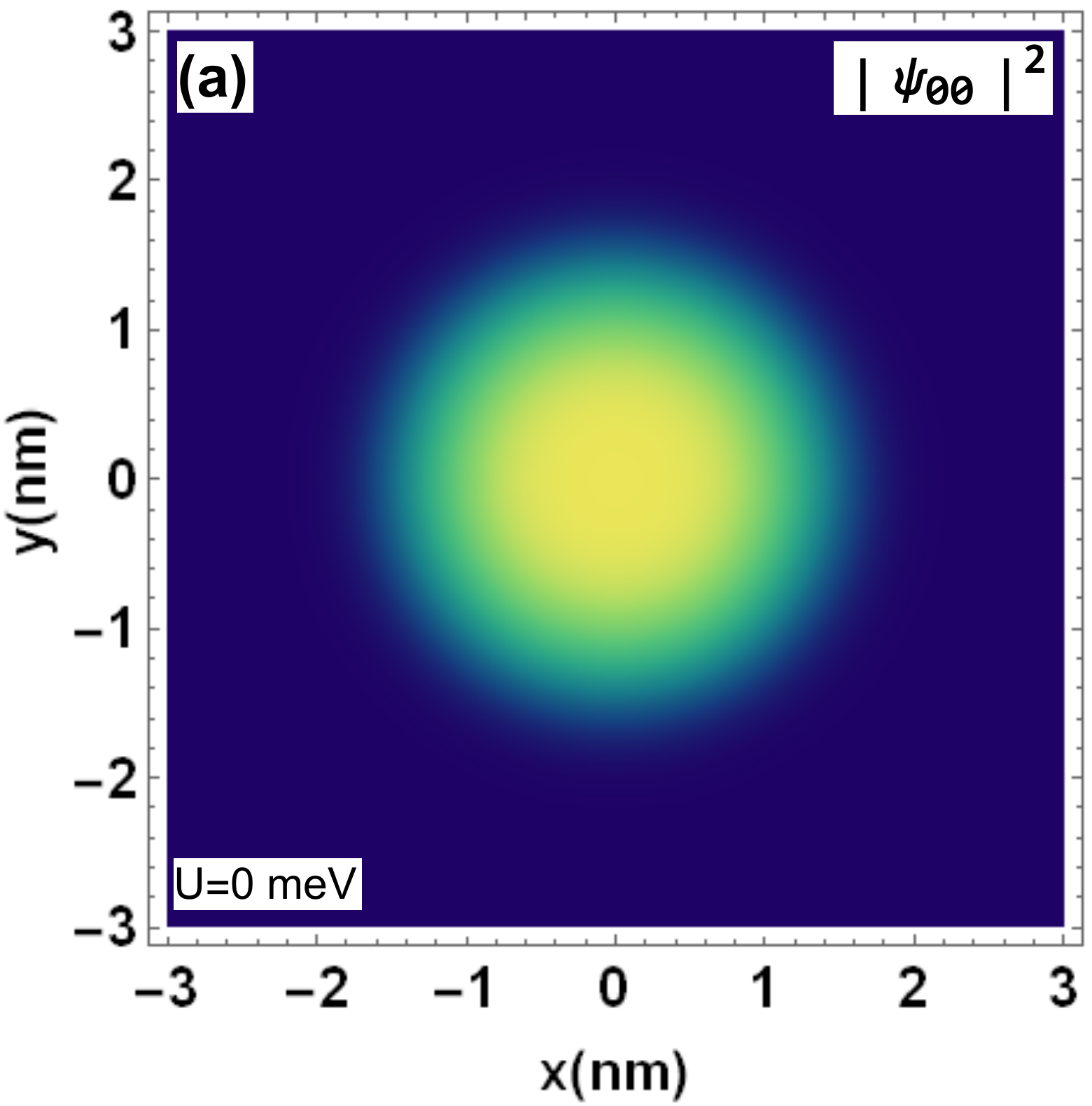} \ \ \ \ \ \includegraphics[width=4.2cm]{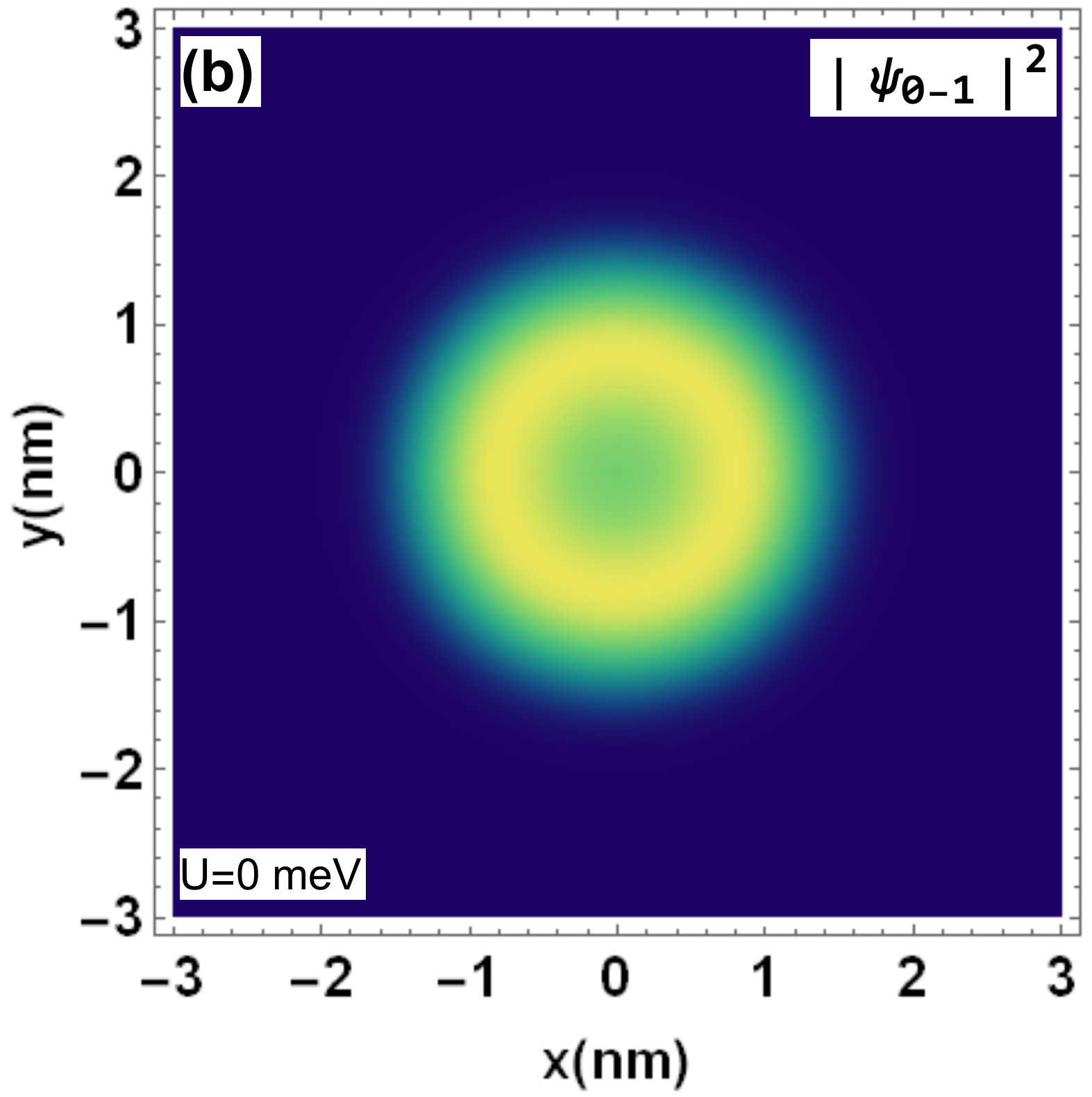}
  \ \ \ \ \ \includegraphics[width=4.7cm]{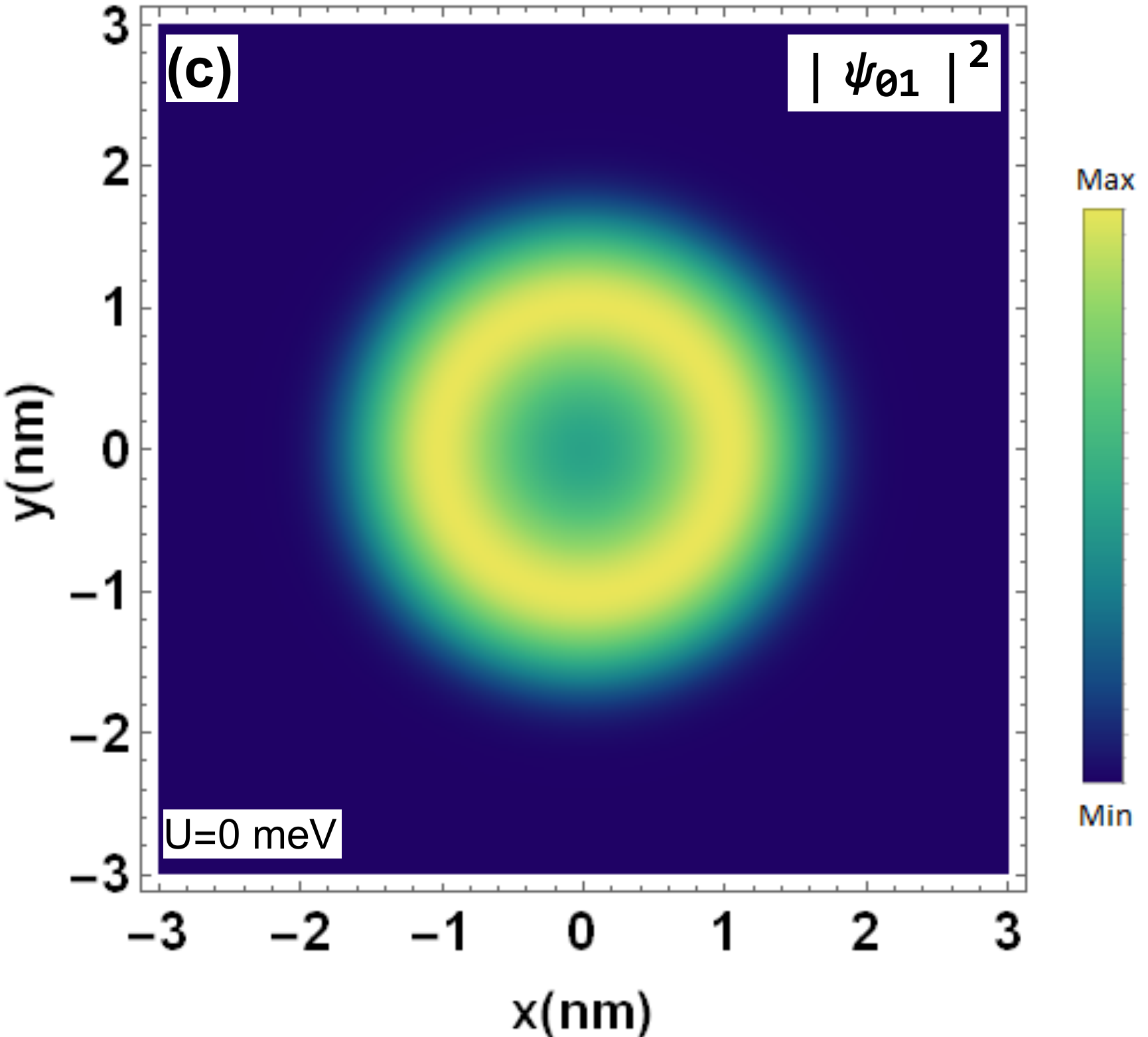}
  \includegraphics[width=4.2cm]{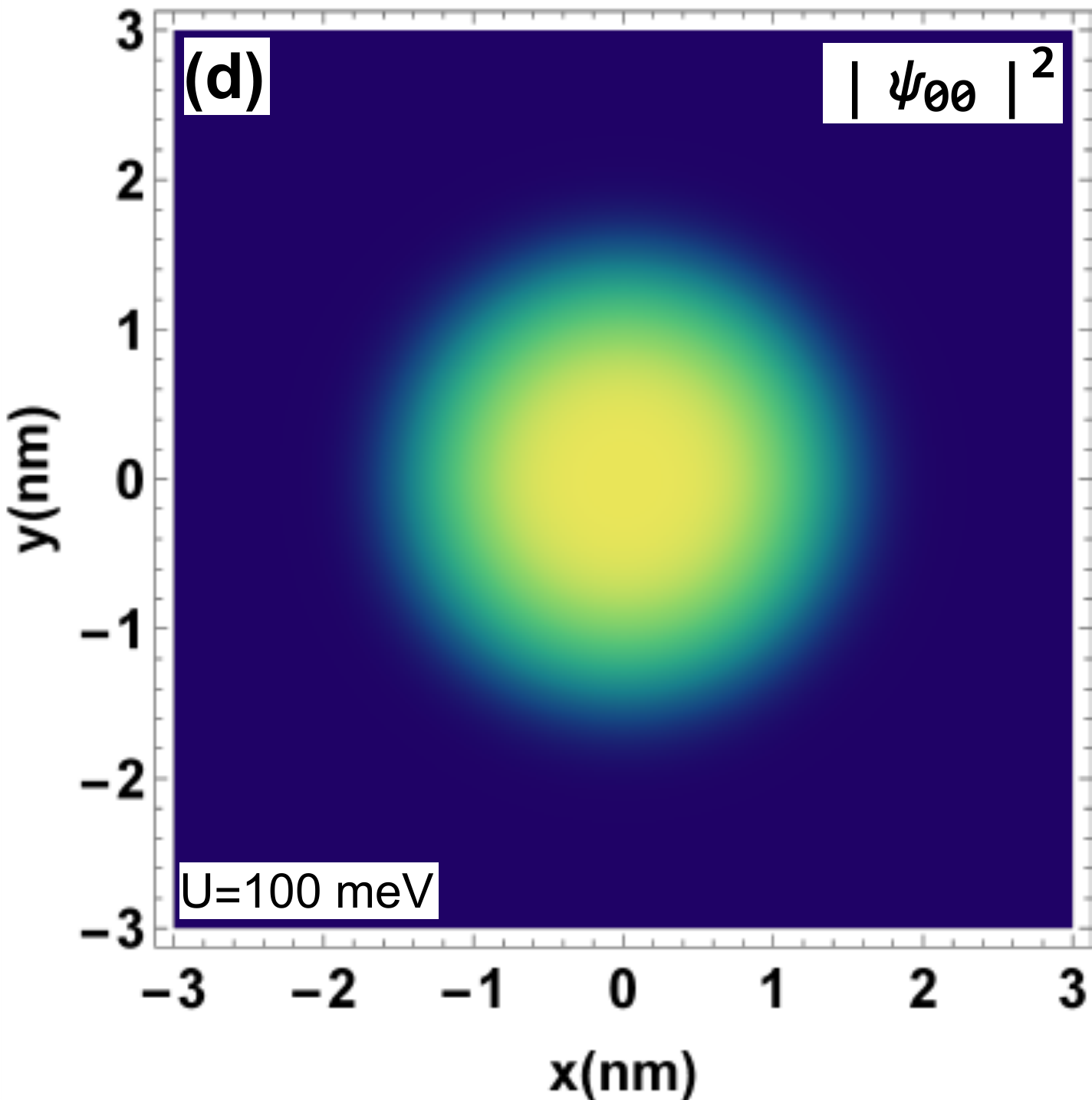}\  \ \ \ \ \includegraphics[width=4.2cm]{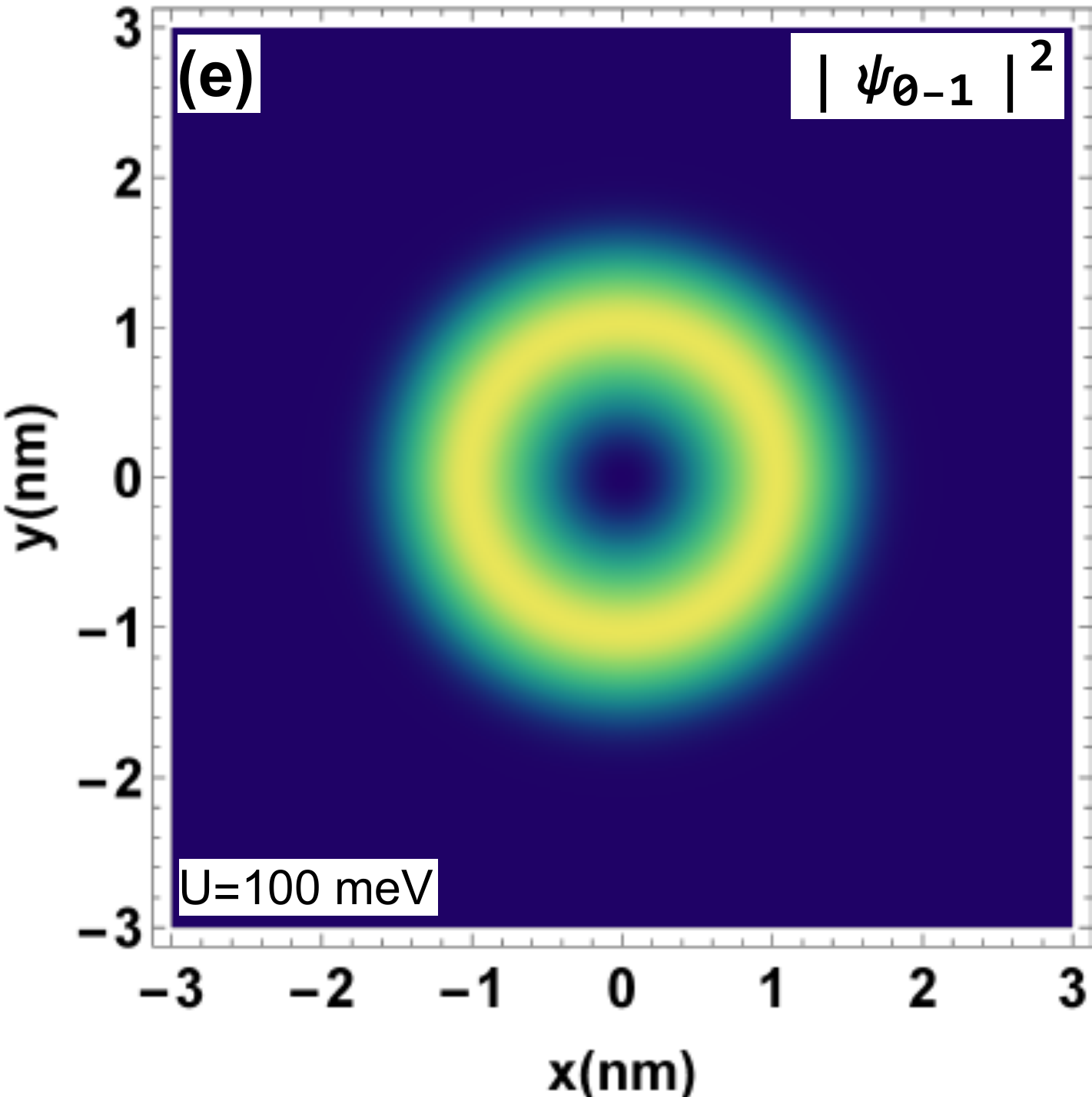}\ \ \ \ \ \includegraphics[width=4.7cm]{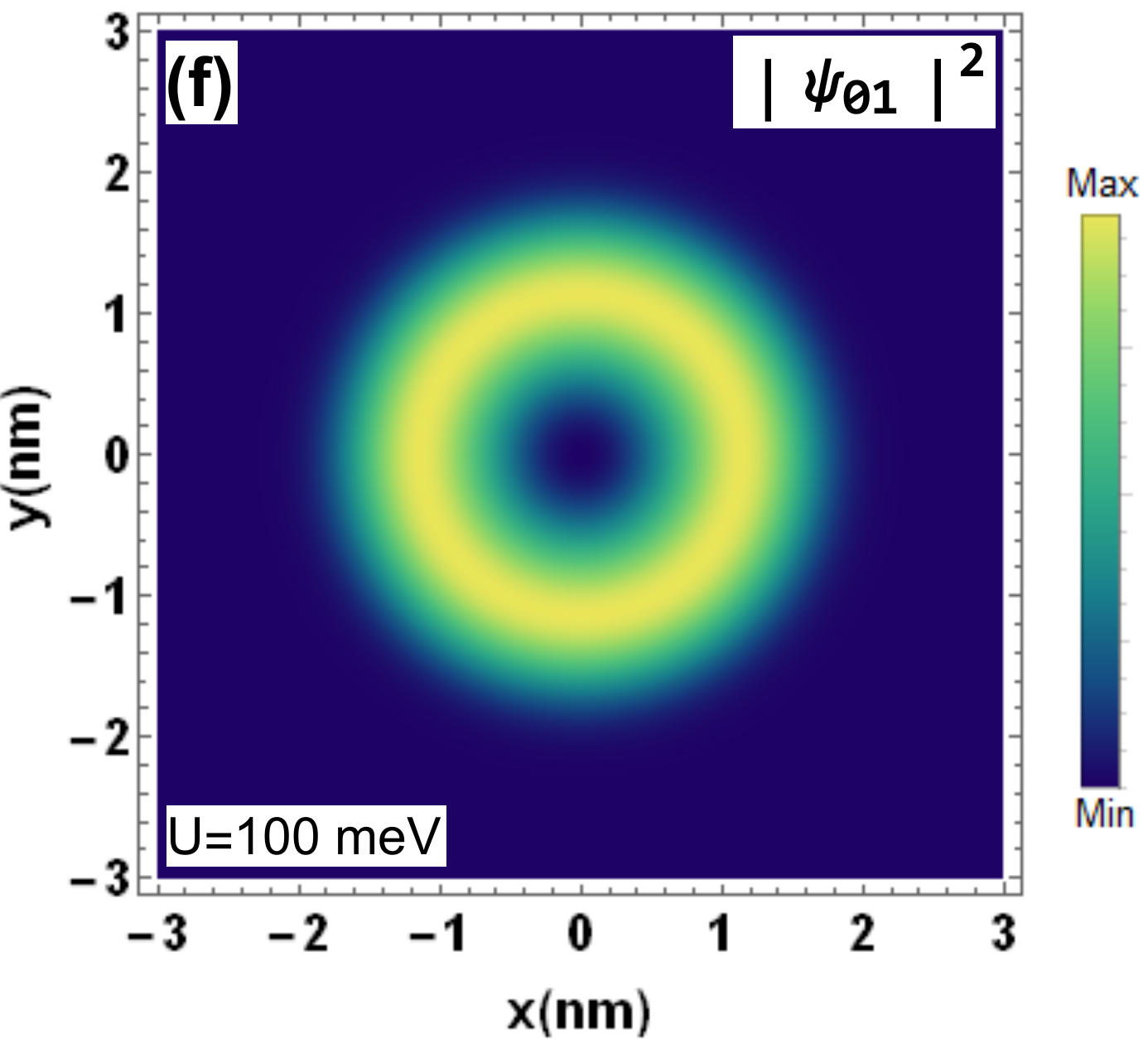}
  \includegraphics[width=4.2cm]{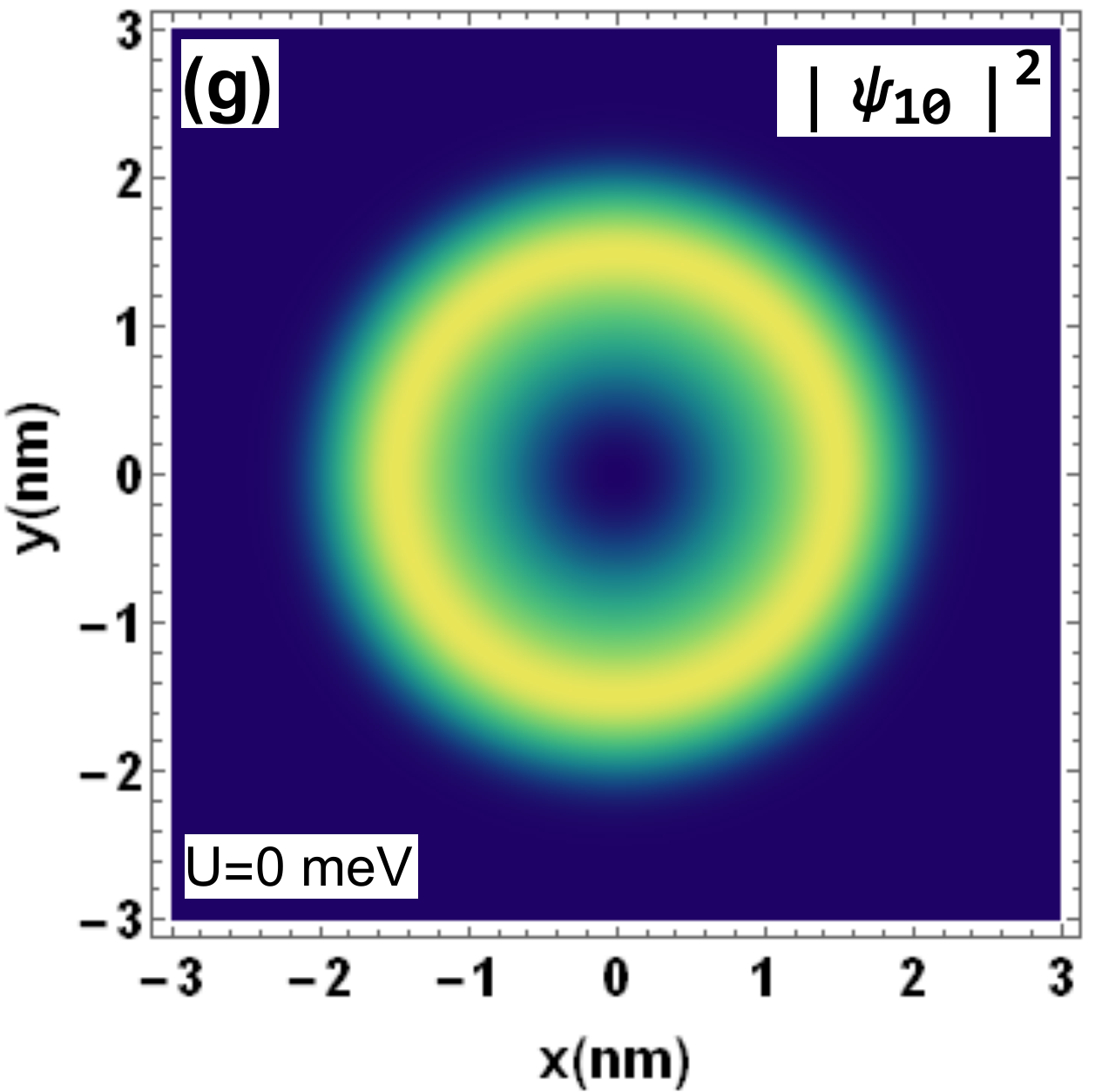} \ \ \ \ \ \includegraphics[width=4.2cm]{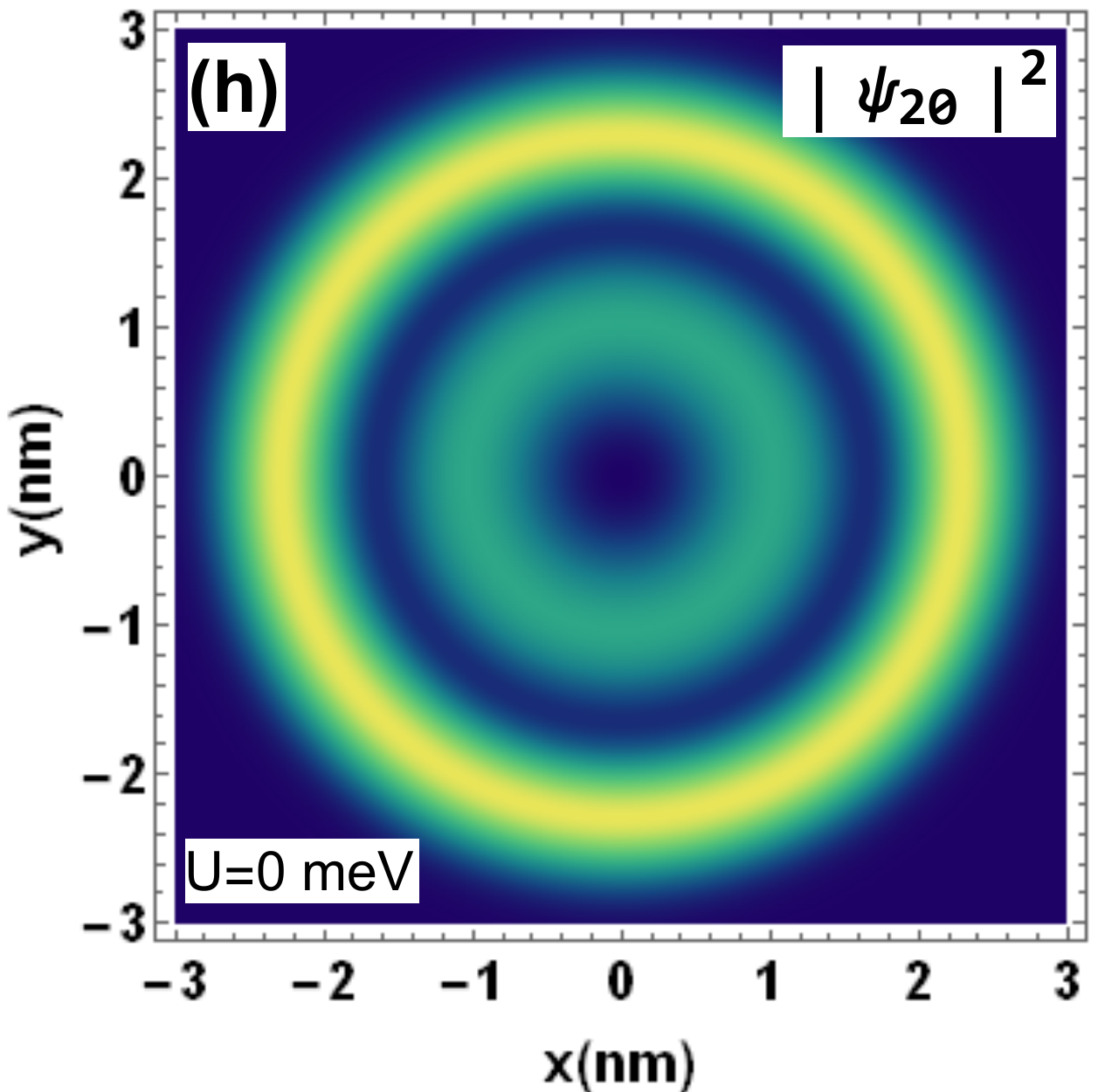}\ \ \ \ \includegraphics[width=4.7cm]{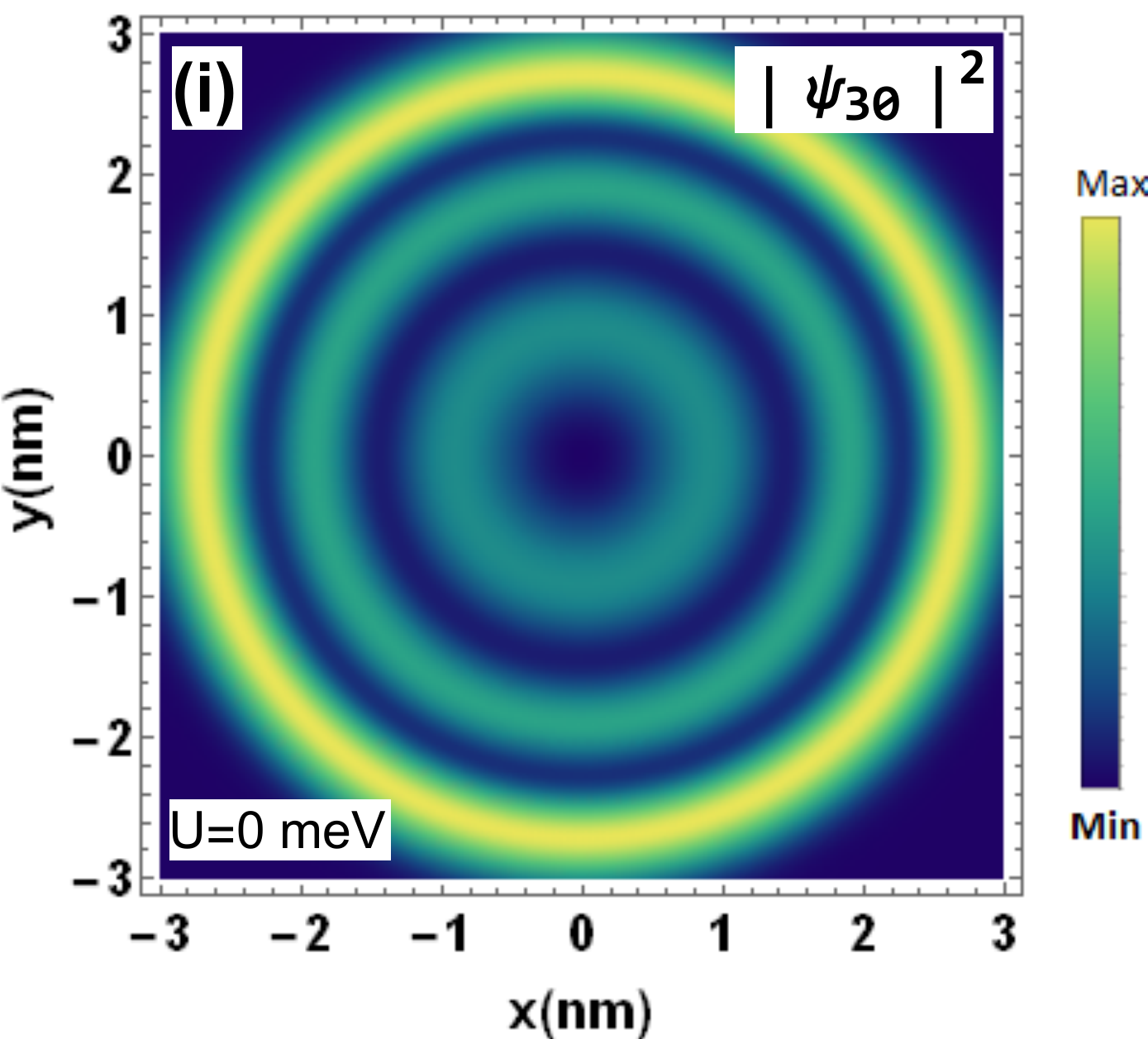}
  \includegraphics[width=4.2cm]{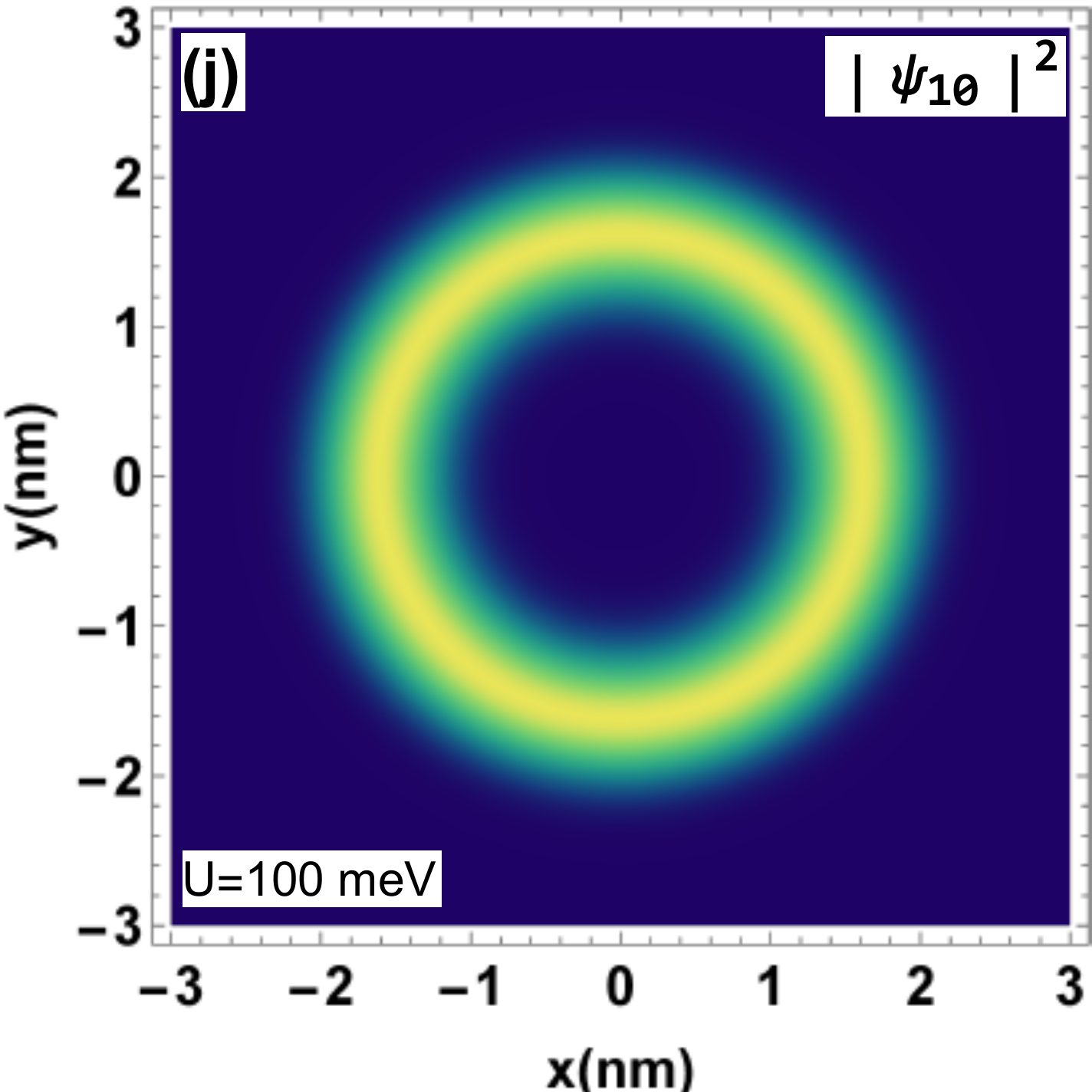} \ \ \ \ \
  \includegraphics[width=4.2cm]{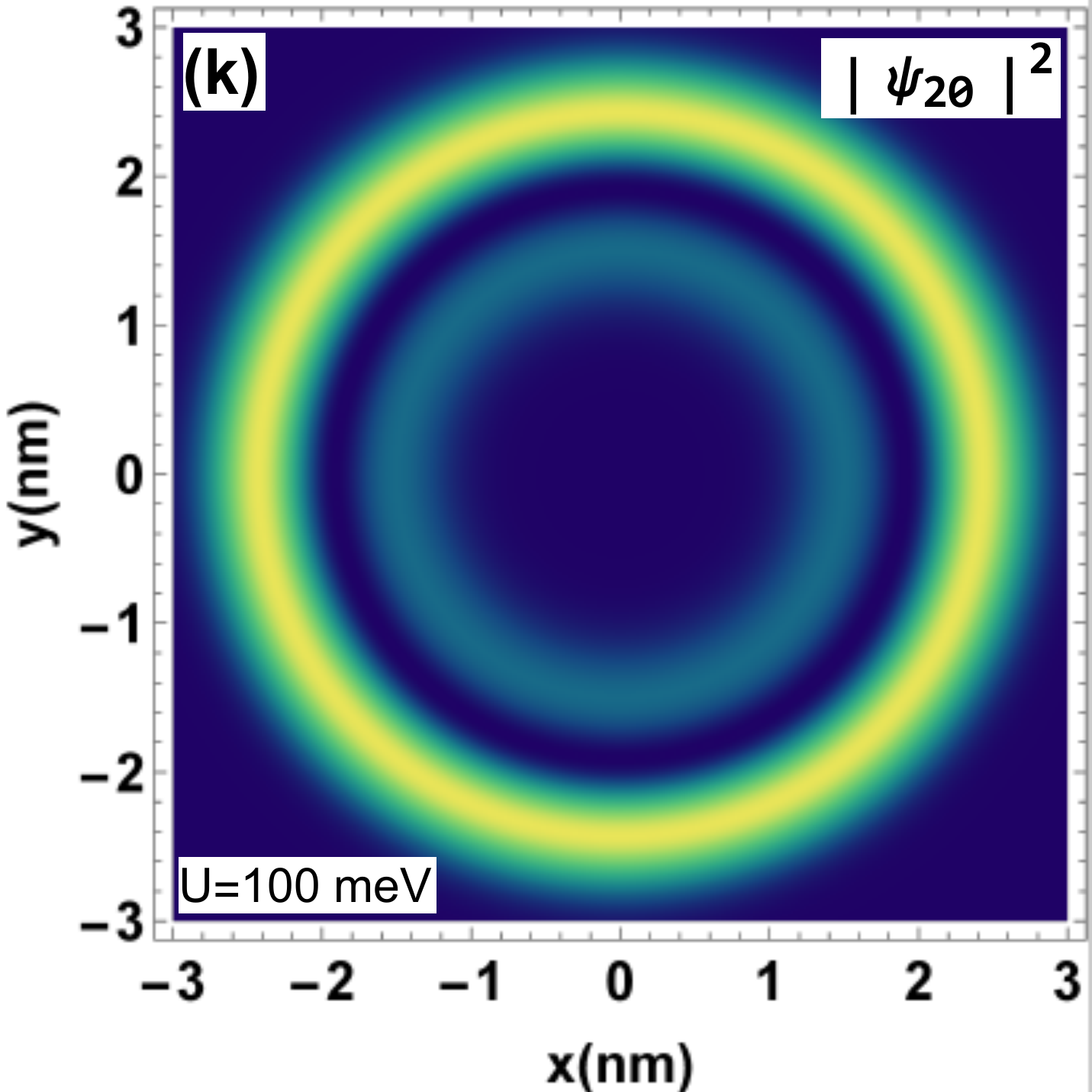}\ \ \ \ \
  \includegraphics[width=4.7cm]{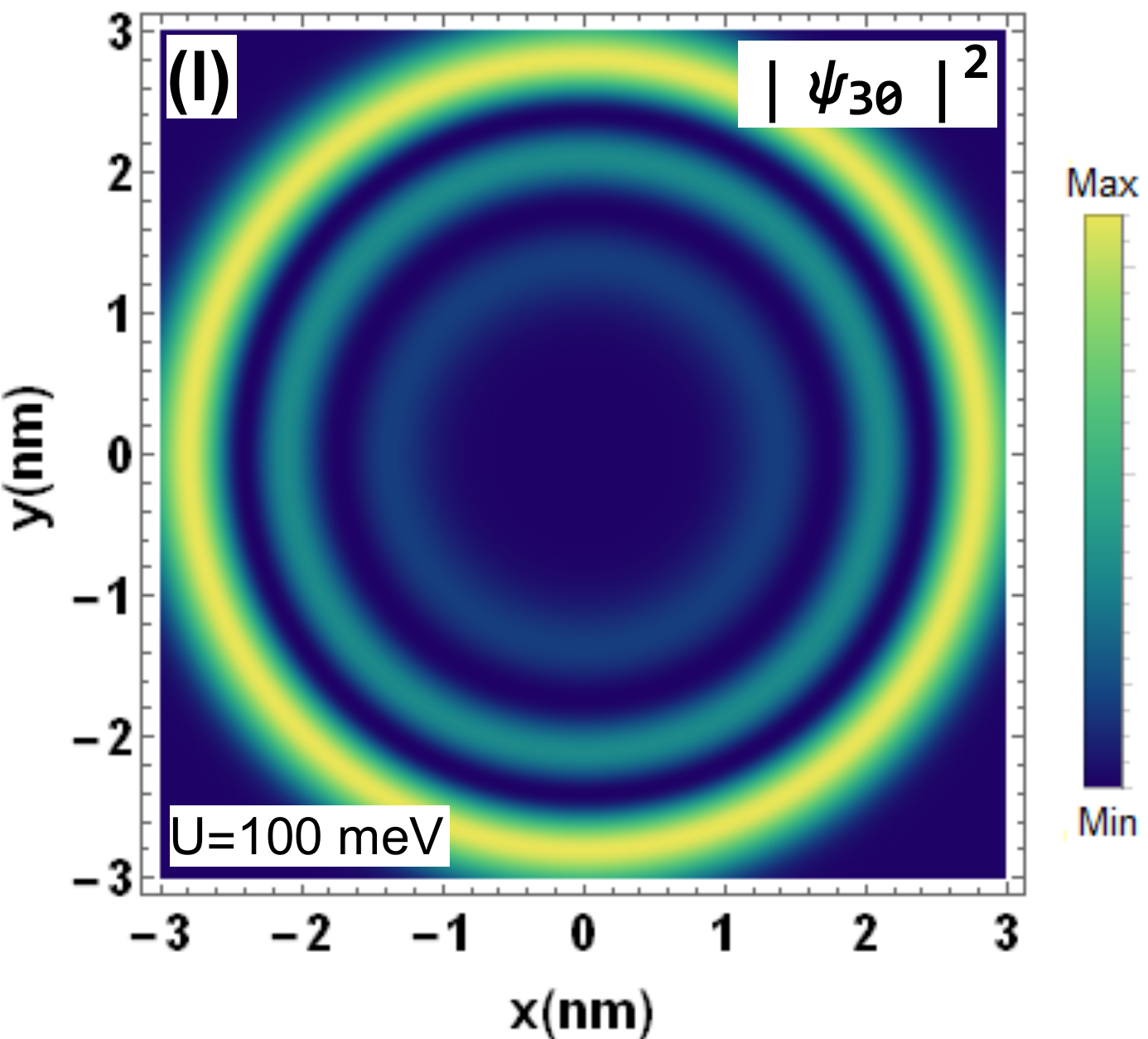}\\
  \caption{(color online) \sf Spatial density $|\psi_{nm}|^{2}$ in the vicinity of the quantum dot. $(a)$: ($n=0$, $m=0$), $(b)$: ($n=0$, $m=-1$), $(c)$: ($n=0$, $m=1$), $(g)$: ($n=1$, $m=0$), $(h)$: ($n=2$, $m=0$), $(i)$: ($n=3$, $m=0$) for $U=0$ meV. $(d)$: ($n=0$, $m=0$), $(e)$: ($n=0$, $m=-1$), $(f)$: ($n=0$, $m=1$), $(j)$: ($n=1$, $m=0$), $(k)$: ($n=2$, $m=0$), $(l)$: ($n=3$, $m=0$) for $U=100$ meV. In all {the} panels $B=10$ T, $R=70$ nm.\label{f6}}
\end{figure}

Figure \ref{f6} shows the electron density $|\psi_{nm}|^{2}$
of the charge carriers in
graphene magnetic quantum dot of radius
$R=70$ nm and subject to the field $B=10$ T for
the potential $U=0$ {meV} and $U=100$ meV,
with some selected values of the quantum numbers $n$ and $m$.
We observe that the electron density for $m=0$ and $n=0$ exhibits a maximum at the center of QD in panels (a,d), while in {the}
panels (g,h,i,j,k,l) for 
$m=0$ and $n=1, 2, 3$ a minimum is observed at the center of the QD.
Note that the low contribution of the other modes of the point leads to the slight asymmetry of the  electron density.
To analyze the effect of the potential $U$ on $|\psi_{nm}|^{2}$,
we presents  {the} panels (g,h,i,j,k)  for a given value $U=100$ meV.
Indeed, by comparison we notice that
for $U=0$ meV  the density is substantial at the center as shown in panels (b,c) but for $U=100$ meV
it has a minima in the center of QD see panels (e,f).
Also the density $|\psi_{n0}|^{2}$ possesses different behaviors in the absence (panels (g,h,i)) and presence
(panels (j,k,l)) of the potential. It is clearly seen that
$|\psi_{00}|^{2}$ has the same behavior with and  without 
 the potential $U$. The interesting result is that the electron density inside the quantum dot is dramatically increased, which is a sign of temporary particle trapping.

\section{Conclusion}

We have studied the confinement of charge carriers in a graphene
 quantum dot submitted to {a} perpendicular magnetic field and  electrostatic potential. Solving the two-band Dirac–Weyl Hamiltonian, in the vicinity of both $K$ and $K'$ valleys, we have obtained analytically the  eigenspinors.
 The boundary conditions were used to
 obtain an equation describing the energy levels
 in terms of the physical parameters characterizing our quantum dot
 {and the strength of the applied magnetic field.}

We have  discussed our results numerically for various choices of the  physical parameters. Indeed, the dependence of  {the} spectrum on {the} radius $R$ of the QD has been investigated 
{the energy spectrum} has the asymmetry $E(m,\tau)=E(m,-\tau)$ for all the values of $\tau$ when {the} limit $R\longrightarrow 0$ is satisfied. As long as $R$ increases, two sets of energy levels appear, one satisfies the symmetry
$E(m,\tau)=E(m,-\tau)$ and the other is not symmetric $E (m,\tau)\neq E(m,-\tau)$.  An energy gap for $m\neq 0$ and at zero energy for $m=0$ has been obtained. 
{According to Figure \ref{f2}, we have observed
that the energy band gap decreases
when the size of the QD increases.}
As far as the magnetic field dependence is concerned,
we have shown that the degeneracy of the valley exists when the magnetic field $B\longrightarrow 0$. By increasing $B$, we have seen that the spectrum exhibits 
anti-crossings and the levels merge at the Landau levels for monolayer graphene, which indicates that the carriers become strongly localized at the center of the graphene magnetic quantum dot.
Furthermore the influence of the electrostatic potential $U$ makes it possible to shift the energy levels vertically by $U$, such that  $E(m,\tau,B,U)=E(m,B,\tau)+U$. Finally, 
 {we} have shown that for some values of the quantum number $n$ and $m$, the electron density in QD is strongly increased, which indicates a temporary electron trapping.

\section*{Acknowledgments}

The generous support provided by the Saudi Center for Theoretical
Physics (SCTP) is highly appreciated by all authors.
AJ and HB
acknowledge the support of KFUPM under research group project RG181001.


\begin{thebibliography}{99}
\bibitem{Novoselov04} K. S. Novoselov, A. K. Geim, S. V. Morozov, D. Jiang,Y. Zhang, S. V. Dubonos, I. V. Grigorieva, and A. A. Firsov, Science 306, 666 (2004).
\bibitem{Novoselov06} K. S. Novoselov, E. McCann, S. V. Morozov, V. I. Fal’ko, M. I.
Katsnelson, U. Zeitler, D. Jiang, F. Schedin, and A. K. Geim, Nat. Phys. 2, 177 (2006).
\bibitem{Zhang05} Y. Zhang,Y. W. Tan, H. L. Stormer, and P. Kim, Nature 438, 201 (2005).
\bibitem{Chung10} K. Chung, C. H. Lee, and G. C. Yi, Science 330, 655 (2010).
\bibitem{Choe10} G. Jo, M. Choe, C. Y. Cho, J. H. Kim, W. Park, S. Lee, W. K.
Hong, T. W. Kim, S. J. Park, B. H. Hong, Y. H. Kahng, and T.
Lee, Nanotechnology 21, 175201 (2010).
\bibitem{Berger06} C. Berger, Z. Song, X. Li, X. Wu, N. Brown, C. Naud, D. Mayou,
T. Li, J. Hass, A. N. Marchenkov, E. H. Conrad, P. N. First, and
W. A. de Heer, Science 312, 1191 (2006).
\bibitem{Ohta06} T. Ohta, A. Bostwick, T. Seyller, K. Horn, and E. Rotenberg,
Science 313, 951 (2006).
\bibitem{Kim09} K. S. Kim, Y. Zhao, H. Jang, S. Y. Lee, J. M. Kim, K. S. Kim, J. H. Ahn, P. Kim, J. Y. Choi, and B. H. Hong, Nature 457, 706 (2009).
\bibitem{Reina08} A. Reina, X. Jia, J. Ho, D. Nezich, H. Son, V. Bulovic, M. S.
Dresselhaus, and J. Kong, Nano Lett. 9, 30 (2008).
\bibitem{De Arco09} L. G. De Arco, Y. Zhang, A. Kumar, and C. Zhou, IEEE Trans.
Nanotechnol. 8, 135 (2009).
\bibitem{Castro09} A. H. Castro Neto, F. Guinea, N. M. R. Peres, K. S. Novoselov, and A. R. Geim, Rev. Mod. Phys. 81, 109 (2009).
\bibitem{Geim09} A. K. Geim, Science 324, 1530 (2009).

\bibitem{Geim07} A. K. Geim and K. S. Novoselov, Nature Materials 6, 183 (2007).


\bibitem{Bacon13} M. Bacon, S. J. Bradley, and T. Nann, Part. Part. Syst. Charact 31, 415 (2013).

\bibitem{Trauzettel07} B. Trauzettel, D. V. Bulaev, D. Loss, and G. Burkard, Nat.
Phys. 3, 192 (2007).

\bibitem{Sun13} H. Sun, L. Wu, W. Wei, and X. Qu, Mater. Today 16, 433 (2013).

\bibitem{Rozhkov11} A. Rozhkov, G. Giavaras, Y. P. Bliokh, V. Freilikher, and F. Nori,
Phys. Rep. 503, 77 (2011).

\bibitem{Katsnelson06} M. I. Katsnelson, K. S. Novoselov, and A. K. Geim, Nat. Phys.
2, 620 (2006).

\bibitem{Espinosa13} T. Espinosa-Ortega, I. A. Luk’yanchuk, and Y. G. Rubo, Phys.
Rev. B 87, 205434 (2013).
\bibitem{Martino07} A. D. Martino, L. DellAnna, and R. Egger, Phys. Rev. Lett. 98,
066802 (2007).

\bibitem{Mirzakhani16} M. Mirzakhani, M. Zarenia, S. A. Ketabi, D. R. da Costa, and
F. M. Peeters, Phys. Rev. B 93, 165410 (2016).

\bibitem{Zebrowski13} D. P. Zebrowski, E. Wach, and B. Szafran, Phys. Rev. B 88,
165405 (2013).


\bibitem{Recher09} P. Recher, J. Nilsson, G. Burkard, and B. Trauzettel, Phys. Rev.
B 79, 085407 (2009).

\bibitem{Liu17} Y. Liu, X. Liu, Y. Zhang, Q. Xia, and J. He, Nanotechnology
28, 235303 (2017).

{
\bibitem{Thomsen}
M. R. Thomsen and T. G. Pedersen, Phys. Rev. B 95, 235427 (2017).

\bibitem{Freitag2}
Nils M. Freitag, Larisa A. Chizhova,Peter Nemes-Incze, Colin R. Woods, Roman V. Gorbachev,
Yang Cao, Andre K. Geim, Kostya S. Novoselov, Joachim Burgdö
rfer, Florian Libisch,
and Markus Morgenstern,
Nano Lett.  16, 5798 (2016).

\bibitem{Tamandani}  Shahryar Tamandani, Ghafar Darvish, and Rahim Faez, 
Appl. Phys. A 122,  37  (2016).

\bibitem{group}
A. Rycerz, J. Tworzydło, and C. W. J. Beenakker, Nature Physics 3, 172 (2007); 
B. M. Hunt, J. I. A. Li, A. A. Zibrov, L. Wang, T. Taniguchi, K. Watanabe, J. Hone, C. R. Dean, M. Zaletel, R. C. Ashoori, and  A. F. Young, Nature Communications 8, 948 (2017). 
}

\bibitem{Recher07} P. Recher, B. Trauzettel, A. Rycerz, Y. M. Blanter, C. W. J.
Beenakker, and A. F. Morpurgo, Phys. Rev. B 76, 235404 (2007).

\bibitem{Heung98} Heung-Sun Sim, K. H. Ahn, K. J. Chang, G. Ihm, N. Kim, and S. J. Lee, Phys. Rev. Lett 80, 7 (1998).

\bibitem{Schnez08} S. Schnez, K. Ensslin, M. Sigrist, and T. Ihn, Phys. Rev. B 78, 195427 (2008).

\bibitem{Abramowitz} Abramowitz M and Stegun I A, Handbook of Mathematical Functions (Dover Publications, Inc, New York, 1965).

\bibitem{Recher009} P. Recher and J. Nilsson, Phys. Rev. B 79, 085407 (2009).

\end{thebibliography}
\end{document}